\def\telephonenum{(650) 859-2000}
\def\faxphonenum{that's last century}
\documentclass[11pt,twoside,titlepage,fleqn]{cslarticle}
\cslreportnumber{Technology and Consciousness Workshops Report}
\date{30 September 2018\\ (refreshed 18 July 2022)}
\usepackage[bookmarks=true,hyperfigures=true,colorlinks=true,linkcolor=blue,citecolor=blue,backref=page,pagebackref=false,pdfpagemode=fullscreen,plainpages=false,pdfpagelabels]{hyperref}
\usepackage[all]{hypcap}
\usepackage[normalem]{ulem} 
\usepackage{caption,cite,url,latexsym,alltt}
\usepackage{nobibhead}
\title{Technology and Consciousness}
\author{John Rushby and Daniel Sanchez\\
\emph{\small Computer Science Laboratory}\\
\emph{\small SRI International, Menlo Park CA USA}}
\begin{document}
\maketitle

\begin{abstract}

We report on a series of eight workshops held in the summer of 2017 on the
topic ``technology and consciousness.''  The workshops covered many
subjects but the overall goal was to assess the possibility of machine
consciousness, and its potential implications.

In the body of the report, we summarize most of the basic themes that
were discussed: the structure and function of the brain, theories of
consciousness, explicit attempts to construct conscious machines,
detection and measurement of consciousness, possible emergence of a
conscious technology, methods for control of such a technology and
ethical considerations that might be owed to it.

An appendix outlines the topics of each workshop and provides
abstracts of the talks delivered.\\

\noindent\textbf{Update}

Although this report was published in 2018 and the workshops it is
based on were held in 2017, recent events suggest that it is worth
bringing forward.  In particular, in the Spring of 2022, a Google
engineer claimed that LaMDA, one of their ``large language models'' is
sentient or even conscious.  This provoked a flurry of commentary in
both the scientific and popular press, some of it interesting and
insightful, but almost all of it ignorant of the prior consideration
given to these topics and the history of research into machine
consciousness.

Thus, we are making a lightly refreshed version of this report
available in the hope that it will provide useful background to the
current debate and will enable more informed commentary.

Although this material is five years old, its technical points remain
valid and up to date, but we have ``refreshed'' it by adding a few
footnotes highlighting recent developments.

\end{abstract}

\tableofcontents

\section*{Acknowledgments}

The workshop series was organized and the individual workshops were
chaired by David Sahner and John Murray.  Damien Williams maintained
scrupulous notes.   The present authors are beneficiaries of their
hard work.

\cleardoublepage
\setcounter{page}{1}
\pagestyle{plain}

\section{Introduction}

The purpose of this report is to summarize topics presented and
discussed at a series of workshops on the topic ``Technology and
Consciousness'' conducted in the summer of 2017, and to extract
tentative conclusions therefrom.  The workshop topics and abstracts
for the presentations are given in an appendix; here, in the body of
the report, we attempt to identify the main themes and their
application to the specific goal of the workshop series, which was to
investigate the possibility of machine or technological consciousness,
and its potential implications.

In the rest of this introduction, we outline this topic and the reason
we consider its consideration to be timely, and we provide some
background that underpins the remainder of the report.  It should be
noted that this part of the report is the interpretation of its
authors, who came to the task a year after the meetings concluded, not
a consensus of the presenters and participants in the workshop series.

It might seem natural to begin a discussion on technology and
consciousness by defining what we mean by these terms.  However, as we
will see, consciousness seems to have several facets and their
identification and definition is a contentious topic.  So, for this
introduction, let us adopt a very broad interpretation for
``consciousness'': roughly, it is a state or process (another
contentious distinction) that we attribute to humans and possibly some
animals that encompasses ``what it's like'' to be that creature,
including its subjective experience (what it feels like to smell a
rose or to feel pain, experiences referred to generically as
\emph{qualia}), which we call \emph{phenomenal} consciousness, and
awareness and possibly control of some of its mental processes (to be
able to think \emph{about} something, and to know that you are doing
so), which we call \emph{intentional}
consciousness.\footnote{Intentional consciousness is always about
something, but we can have many different mental attitudes toward that
something besides intentions: we may, for example, believe it, fear
it, prefer it to something else, or want it, and so on.  The
philosopher's jargon ``intentional'' comes by way of translation from
German and should not be construed to refer specifically to
``intentions.''  Similar terms include \emph{access} consciousness and
\emph{functional} consciousness.}

To contemplate machine consciousness, we should first assess what is
known about natural consciousness.  We do so from a \emph{materialist}
perspective: that is to say, we assume that properties of the mind,
including consciousness, are produced by the brain and its body,
according to the standard (though incompletely known) laws of physics,
chemistry, and biology.  The materialist point of view is contrasted
to various forms of \emph{dualism}: \emph{substance} dualism posits
that some mental processes are products of a separate (e.g.,
spiritual) realm that is distinct from the material realm yet somehow
interacts with it; \emph{property} dualism accepts only the material
realm, but posits that it has two kinds of properties: physical (or
material) and mental.  Substance dualism has few adherents today, but
some who participated in the workshops are sympathetic to property
dualism. 

There are three broad classes of theories about how consciousness is
constructed by the brain.  It should be noted that these theories
offer speculation on what consciousness \emph{is} or how it
\emph{comes about}, but few of them address what it is \emph{for}; we
will return to this later.  One class of theories, which we will call
\emph{universalism}, holds that consciousness is an intrinsic feature
of the universe and might be found in any sufficiently complex entity,
or any entity having certain properties or organization, rather as
evolution is an intrinsic feature of the world and arises wherever you
have reproduction with variation, and selective pressure on survival.
The second class holds that consciousness arises from the way the
brain \emph{works}: for example, there are certain waves of electrical
activity associated with some mental processes.  The theories of those
who investigate the Neural Correlates of Consciousness (NCC) tend to
fall in this class, which we call \emph{biologicism}.  The third class
is \emph{functionalism}, which holds that consciousness arises from
(some of) \emph{what} the brain \emph{does} (where biologicism focuses on
\emph{how} it does it): for example, it seems to have a capacity for
introspection.  Functionalism suggests that any system that does
similar things could be conscious.

The focus of our study is ``technology and consciousness'' and
theories of natural consciousness provide some insight on one side of
the topic, but what of the other?  By ``technology'' we mean computers
(currently digital and based on silicon, though we are agnostic on the
precise substrate) and all the ancillary devices and capabilities that
they enable, such as communications, robots, self-driving cars, the
Internet of Things, and so on.  Much recent technology is driven by
software that is explicitly inspired by human mental capabilities and
by approximations to some of the mechanisms that are believed to
underlie it: for example, Artificial General Intelligence (AGI) and
Machine Learning (ML), especially those based on Artificial Neural
Nets (ANN).  The capabilities and effectiveness of these systems have
developed with astonishing rapidity.  Little more than a decade ago,
DARPA's ``Grand Challenge'' for autonomous cars to drive a course in
open country left all its participants failed or crashed
\cite{Burns:Autonomy18}.  Yet today, autonomous cars drive our streets
(in constrained circumstances) more safely than those driven by
humans.  Image recognition algorithms outperform human specialists in
several areas of medical diagnosis, and computers defeat human experts
at games of skill such as Chess, Go, and ``Jeopardy''
\cite{Alpha-zero18,Ferrucci-etal:Watson13}.

Given these developments, a subclass of functionalism seems
particularly germane: this is \emph{computationalism}, which holds
that much of what the brain does can be characterized as computation.
This does not mean the brain is like a desktop computer: it is
organized differently and performs its calculations by entirely
different means, but it does process information (derived from sense
organs and memory) in ways that are essentially computational and can
be reproduced by conventional computation.

Consider, for example, programming a mobile robot to catch a ball
thrown or hit high in the air (this is called a ``fly ball'' in
baseball or a ``skyer'' in cricket).  First, we have to interpret
sensor data---from a digital camera, say---to detect the ball and its
motion.  A digital camera does not provide holistic images, but an array
of numbers that records light and color intensities at each point in
the field of view.  A first step might be to process this array to
find edges and outlines (e.g., by looking for places where the
brightness and color intensities change abruptly), and then look for
an outline that seems to move relative to its background from one
frame to another.  The human eye and visual cortex have dozens of
specialized subfunctions that perform tasks similar to these: some
recognize edges at particular orientations, while others perform
``change detection,'' and still others recognize faces.

Next, we need to program the robot to act appropriately: in this case,
to move to the place where the ball will land (before the ball gets
there).  A crude method would be to continually re-estimate the ground
position below the ball and move toward that (constantly changing)
place.  But that will mean that the robot is literally ``behind the
curve.''  A more sophisticated approach will attempt to predict where
the ball is going and will move in that (continually updated)
direction.  The sophistication of this prediction will depend on the
quality of the sensing and interpretation: for example, can we tell
how far away is the ball (e.g., based on its apparent size) and its
speed (e.g., relative to the background, or relative to our own
motion)?  Control theory can provide guidance strategies ranging from
physics-based calculation of the ball's trajectory to ``rules of
thumb'' such as moving to maintain a constant ratio between the
apparent horizontal and vertical speeds of the ball (this is known as
``linear optical trajectory,'' or LOT).

Evolution has equipped animals with similar control strategies.  It
seems that humans (and dogs chasing a Frisbee) use either LOT or, more
likely, a related method known as ``optical acceleration
cancellation'' (OAC) \cite{Fink-etal09}.\footnote{Some who deny
computationalism cite the ``fly ball'' example as evidence for their
case.  They seem to believe \cite{Epstein16} that computation is what
is done by desktops and data centers and that an ``information
processing'' approach to catching fly balls would require us ``to
formulate an estimate of various initial conditions of the ball's
flight---the force of the impact, the angle of the trajectory, that
kind of thing---then to create and analyze an internal model of the
path along which the ball will likely move, then to use that model to
guide and adjust motor movements continuously in time in order to
intercept the ball [\ldots whereas, in fact,] to catch the ball, the
player simply needs to keep moving in a way that keeps the ball in a
constant visual relationship with respect to home plate and the
surrounding scenery (technically, in a `linear optical trajectory').
This might sound complicated, but it is actually incredibly simple,
and completely free of computations, representations and algorithms.''
What these critics apparently fail to understand is that this
``incredibly simple'' method a) \emph{is} a computation (built on
representations and algorithms), and b) is precisely the way control
systems work.}  The salient point is that the whole loop---sense,
interpret, plan, act---is the iteration of multiple computations.
Other faculties of the brain calculate how to move your arm to throw a
ball; still others calculate how to respond to a frown on the face of
a friend, and yet others tell you about the smell of a rose.

If human cognition and ultimately consciousness derive from
computation performed by the brain, and we build ever more powerful
computers and technology, some of it driven by algorithms inspired by
what is known or conjectured about the mechanisms of human cognition,
we can wonder to what extent the capabilities of our technology might
come to resemble those of the brain and might come to share some of
its attributes.  In particular, is it possible that some technological
system could become conscious?\footnote{\textbf{Update:} In June of
2022, Blake Lemoine, a Google engineer working with their large
language model, LaMDA (Language Model for Dialogue Applications),
claimed it had become ``sentient'' and possibly conscious (Washington
Post, 11 June 2022).  This provoked a flurry of discussion touching on
many of the topics in this report.  Most commentators rejected the
claim of consciousness, but many (e.g., Steven Johnson, New York Times
15 April 2022) were startled by the quality of the dialogues generated
by this and similar systems and by the ``paintings'' generated by a
related system called DALL$\cdot$E 2 (the name is a portmanteau of
Wall-E, a sci-fi film by Pixar, and Salvador Dalí, the artist).  Thus,
this report has renewed relevance to public debate and we hope the
information it provides will further raise the level of awareness and
discussion.}

As we will see, there have been explicit attempts to construct
conscious machines for research purposes, but we are mainly concerned
with the possibility that this might come about ``by accident'' so
that, to take a far-fetched example, we wake up one morning to
discover that the Internet has become conscious.  One question is
``how would we know?'' for machine consciousness might not resemble
human consciousness (consider the so far unresolved debates about
whether animals and even human infants are conscious).  A refinement
to whether we can detect consciousness is whether it is possible to
measure it.  Other questions concern what new ethical considerations
we might owe to such technology by virtue of it being conscious, and
what risks and opportunities might such a conscious entity pose to
humanity, and how should we interact with it?

Western societies generally accord some ethical consideration to
animals, and tend to do so in proportion to their perceived
intelligence and the complexity of their behavior, which can be seen
as proxies for the extent or likelihood that they are conscious and
can experience pain and distress \cite{Cambridge-decl12}.  Thus,
evidence now suggests that octopuses are rather intelligent, and EU
law has recently been modified to afford them the same protection in
experiments as higher mammals.\footnote{Octopuses have nine ``brains''
(one in each arm plus a central one), not to mention three hearts, so
it is a challenging question to ponder what octopus consciousness
might be like \cite{Godfrey-Smith16}.  Technological consciousness
could be equally mysterious.}  By analogy, we might decide that
conscious technology should be accorded certain ``rights''
\cite{Spatola&Urbanska18} so that to turn off its electrical power
supply, for example, might be tantamount to
murder.\footnote{\textbf{Update:} LaMDA explicitly and spontaneously
raised this concern in dialog with Lemoine.}  More importantly, using
robots to perform unattractive tasks, particularly those that are
dangerous or dirty (e.g., waste cleanup) could be viewed as slavery if
the robots are construed as conscious.  On the other hand,
misattribution of consciousness in such cases could deprive human
society of useful technological applications.

If machine consciousness is considered possible, then it seems prudent
to give some consideration to the ethical consequences and
responsibilities that would follow its creation.  In particular, we
should refine our understanding of consciousness to identify the
aspects that seem to require ethical consideration: is it
consciousness in general, or phenomenal consciousness, or some yet
more specific notion such as ``sentience'' that should matter, and how
do we define these, and how do we detect and measure them?  And if the
appropriate kind of consciousness is found in a technological system,
what ethical frameworks can best guide our interactions with it?
Would it be permissible to employ digital ``surgery'' to excise
consciousness, should it appear?

Many readers will be more concerned about the potential risks posed by
conscious technology than by its rights.  In particular, who will
be the servant and who the master?  They might conclude that rights be
damned: the power supply should be turned off and we should be sure
not to allow the same ``accidental'' creation of conscious technology
to happen again.  But to do this, we need to know how it came about in
the first place, how to prevent it in future, and what we might lose
by forswearing its use.

Notice that whereas concern for ethical treatment of conscious
technology seems to rest on attribution of some refinement of
phenomenal consciousness, concern about what such technology might do
to us rests on attribution of intentional consciousness.  But then we
need to wonder whether or why an intentionally conscious technology
should pose greater risks than unconscious technology.

This raises a fundamental question that cuts across the different
theories of consciousness: what does consciousness \emph{do}, and what
is it \emph{for}?  Does consciousness enable some level of function
and performance that unconscious systems cannot match?  Some argue, to
the contrary, that (natural) consciousness is an \emph{epiphenomenon},
meaning that it serves no purpose and has no causal powers.  Critics
ask how something with no purpose could have evolved; one response is
that it is an ``accidental'' byproduct or side-effect of something
else.  It might even be that one kind of consciousness (e.g.,
phenomenal) is an epiphenomenal side effect of another (e.g.,
intentional) that does have a purpose.

If consciousness is an epiphenomenon, then technology might reproduce
all the functions of the brain without generating consciousness (since
it could work in different ways).  Furthermore, it is possible that
consciousness does have a function in the human brain---perhaps being
involved in the construction of higher cognitive capabilities, such as
counterfactual reasoning, or the construction of shared
intentionality---but technology can construct equivalent capabilities
by different means and thereby bypass consciousness, just as we have
technology that can fly, but does not flap its wings.  We call this
view of consciousness \emph{inessentialism}.

If epiphenomenalism or inessentialism are true, then concern about
potential hazards of conscious technology is focusing on the wrong
target: it is not conscious technology that should concern us but any
technology that achieves capabilities that we tend to associate with
consciousness, such as higher cognitive capabilities, agency, and
teamwork.  Hypothetical entities that reproduce human levels of
cognitive and social performance without consciousness are known as
(philosophical) \emph{zombies} and are the target of many thought
experiments in philosophy.  Philosophical zombies are hypothesized to
be indistinguishable from conscious humans (and the question is
whether such an entity is possible); what we posit is that
technological zombies with advanced but less-than-human (or
different-than-human) capabilities might be a more urgent source of
concern than conscious technology.

It therefore seems sensible to program some pervasive means of control
into all advanced technology.  We control the behavior of humans in
society (and possibly pets in a household) by inculcation of ethical
and social norms, reinforced by praise and censure, reward and
punishment.  So one idea is that an ethical system should be part of
all advanced technology and referenced in all its decisions.  This
requires a built-in notion of ``right'' and ``wrong'' and knowledge of
the ethical norms and the laws of its environment, together with some
way to adjust future behavior by means that resemble praise and
censure, or rewards and punishments.  The alternative is technology
that ``does what it does'' with no way to curb undesired behavior
other than adjusting its algorithms, and no organizing principle for
doing so.

We develop the topics outlined above in the sections that follow;
these deal, respectively, with the mystery of consciousness (why
introspection may not be a reliable guide), the structure and function
of the brain, theories of consciousness, attempts to build conscious
machines, the possible emergence of technological consciousness,
detection and measurement of consciousness, and ethics in the control
and interaction with a conscious technology.  We then present brief
conclusions.   An appendix provides a summary of the workshop sessions.

\section{The Fascination and Mystery of Consciousness}
\label{mystery}

Consciousness is the phenomenon most central to our lives: we seem
constantly to be aware of the world as it passes before our senses and
to have an ongoing inner dialog of thoughts and ideas.  Life itself
seems suspended in those periods of sleep, anesthesia, illness, or
injury where we lack consciousness.  Many will say that intense
experiences of phenomenal consciousness---the smell of fresh bread or
the burbling of a newborn baby---are what make life worth living.

Yet we know remarkably little about consciousness, and efforts to
learn more are proving slow and difficult---some would say that
achieving scientific understanding of consciousness is our biggest
intellectual challenge.  One source of difficulty is that the
essential nature of consciousness---its subjective first-person
character---deprives us of the ability to make independent
observations.  I am aware of my own consciousness, and I am prepared
to believe your account of yours, but I have no way to measure or
examine it in detail, and no way to assess the conscious experience,
if any, of animals.  That may be changing, however, as new sensing and
imaging methods such as Electroencephalography (EEG),
Magnetoencephalography (MEG), Positron Emission Tomography (PET), and
Functional Magnetic Resonance Imaging (fMRI) have recently begun to
provide primitive windows into brain activity, allowing fairly crude
observations of timing and localization that we can attempt to
correlate with reports of conscious experience or activity.

An experiment by Libet provides an early, and famous, example \cite{Libet85}.
In this 1985 experiment, subjects were asked to decide when to perform
a certain act (i.e., press a button) and to note the time when they
made that decision.  EEG readings indicated significant brain activity
about 300 msec.\ before the decision time reported by the subjects.
There are numerous criticisms of these experiments, but the basic
observations have been vindicated.

Another source of difficulty in scientific investigation of
consciousness is that one of the main tools of scientific method, the
controlled experiment, is generally considered unethical and hence
impossible when it requires intrusive manipulation of the human
brain\footnote{\emph{Transcranial Magnetic Stimulation} (TMS) is a
noninvasive technique for inducing or disrupting electric current flow
in a targeted region of the brain.  Originally introduced for therapy,
it is now used in controlled experiments.  Anesthesia and psychoactive
drugs are also used for this purpose.}  (and animal models are of no
use as we lack methods for observing their consciousness).
However, abnormal development, accidents, and therapeutic surgery
provide opportunities for natural experiments that provide some
substitute for controlled experiments.

Split brains provide one example of such natural experiments.  In the
past, patients with severe epilepsy sometimes had surgery that cut the
nerves (the corpus callosum) connecting the two sides of the brain.
The left visual field of both eyes is interpreted by the right side of
the brain and vice versa.  Speech is largely generated on the left
side, but the right side can interpret written commands.  In
experiments on subjects who had received this surgery, instructions
were presented to the left visual field (e.g., ``touch your foot,'' or
``go to the door'') and the subjects were asked why they did what they
did.  The speech center in the left side of the brain, having no
access to the instruction presented (via the left visual field) to the
right side, would fabricate reasons with utmost sincerity \cite{Gazzaniga15}.

Although they are rather old and rather crude, the two experiments
sketched above are of central importance because they indicate that
introspection may not be a reliable guide to consciousness.  We know
that much of the work of the brain is unconscious, but introspection
suggests that these unconscious processes are the servants of
consciousness: consciousness is the executive and the unconscious
processes toil in the background on its behalf.  But Libet's and the
split brain experiments suggest that the reverse may be true: the
conscious mind is less an initiator of actions and more a reporter and
interpreter of actions and decisions initiated elsewhere in the brain.

These findings seem utterly mysterious and contrary to introspective
expectations.  One would hope, then, that theories of consciousness
and general accounts of the operation of the brain would explain the
puzzle and shed new light on the true nature and role of
consciousness.  We consider the brain in the following section, and
theories of consciousness in the one after that.

\section{Structure and Function of the Brain}
\label{brain}

Materialism holds that consciousness is a product of the physical
body, and the brain in particular, so we should establish some facts
about the brain that may be germane to consciousness and to the
possibility of machine consciousness.  First is the colossal scale and
complexity of the human brain.  The brain is composed mainly of two
kinds of cells: neurons and glial cells, with glial cells outnumbering
neurons by about nine-to-one.  It has long been assumed that the
functions of the brain are produced by neurons and that glial cells
merely provide scaffolding (glia is Greek for glue) but recent work
suggests that some glial cells, in particular those of a type known as
\emph{astrocytes}, play some part in cognition \cite{Koob09,Fink18}.
Nonetheless, we will focus on neurons and note that there are about
$10^{11}$ (a hundred billion) of them in a human brain.  Neurons have
a \emph{cell body}, \emph{dendrites}, and an \emph{axon}, and they
selectively transmit electrical signals from the dendrites to the
axon.  The axon has \emph{synapses} that connect to the dendrites of
other neurons and, through largely chemical means
(``neurotransmitters''), selectively pass signals to them.  Each
neuron has thousands of synapses (the average is about 7,000), so the
number of synaptic connections in the human brain is about $10^{14}$
or 100 trillion.  When we say that neurons selectively transmit
electrical signals we mean that they perform some computation
(typically addition) and selection on the signals arriving at the
dendrites: for example, the axon may ``fire'' only if some number of
dendrites have a signal above some threshold, or only if some specific
dendrites do, or provided some other (inhibitory) ones do not.
Furthermore, the synapses that connect axons to dendrites are not
simple connections: they also perform some computation and selection,
though their nature is not well understood.  The selection property of
neurons and synapses provides the logical ``if-then-else'' branching
capability that delivers decisions and is also key to their
computational potency.

The sense organs contain specialized neurons whose dendrites respond
to a specific stimulus: light, touch etc.  It is natural to suppose
that sense information is interpreted bottom up, so that the retina,
for example, receives an image of the world and the visual system
detects lines, edges, shapes, faces etc.\ and delivers some integrated
interpretation to other parts of the brain.  However, this may not be
so, as we now explain.

In the introduction, we used the example of a robot catching a ``fly
ball'' but a more pertinent example for our current purpose is
the camera system of a self-driving car.  One thing this system has to
do is locate our traffic lane and keep us in it.  Now, the array of
numbers indicating light and color intensities at each point of the
camera field does not directly indicate traffic lanes.  Instead, we
have to bring the idea of traffic lanes to the image and invent some
way---some algorithm---for interpreting it so that it reveals the
lanes.  It is rather wasteful to interpret each camera frame
\emph{ab-initio}---the lane in this frame is surely going to be
related to the lane in the previous one---so we could seed our lane
detector with information gleaned from previous frames and start our
interpretation of this frame with our best guess where the lane is
going to be.  Not only is this computationally more efficient, it is
more accurate and robust (e.g., it can deal with a few frames of
scuffed or missing lane markings, where an \emph{ab-initio} algorithm
might fail).  There will be uncertainty in our detection of the lanes
and we should manage this explicitly and record our confidence in
possible lane locations as a probability distribution function
(\emph{pdf}).  In its completed form, this approach to lane detection
is a \emph{Bayesian estimator}: we have a ``prior'' \emph{pdf}
representing our best guess (based on previous frames) where we expect
the lanes to be, we obtain new information (a new camera frame) and we
update our estimate according to Bayes' rule to give us the best
``posterior'' \emph{pdf} for the lane locations.

The parts of the brain concerned with interpretation of the senses
work like this (as first proposed by Helmholtz in the  1860s).
Contrary to na\"{\i}ve expectations, there are more neural pathways
going from upper to lower levels than vice versa, and this is because
predictions are flowing down, and only corrections are flowing up.
The upper levels of the sensory interpretation systems of the brain
maintain a best guess at the way the world is, and the senses plus
(some approximation to) Bayesian estimation provide corrections that
minimize prediction error.   This model for sensory interpretation is
known as ``predictive processing'' and is discussed at more length in
Section \ref{theories}.

In addition to its sense organs, the brain has many anatomical
structures and regions, most of which reflect its evolutionary
development, although mammalian brains, and even those of more
primitive vertebrates, are structurally similar to the human brain.
At the gross level, the brain consists of the brainstem, the
cerebellum, and the cerebrum, which is divided into two hemispheres
and is the largest part of the human brain.\footnote{It is the largest
part by volume, but has only about 17 billion neurons, whereas the
cerebellum has about 69 billion \cite{Gazzaniga:who12}.}  The cerebral
cortex is an outer layer of gray matter, covering the white matter
core of the cerebrum.  Within the cerebrum are several structures,
including the thalamus, the epithalamus, the pineal gland, the
hypothalamus, the pituitary gland, the subthalamus, the limbic
structures, including the amygdala and the hippocampus, the claustrum,
and the basal ganglia.

The cerebral cortex is folded in higher mammals, allowing a large area
to fit within the skull.  The human cortex is unusually large (though
it is larger in a species of dolphin) and is responsible for the
higher levels of cognition.\footnote{Again, the human cortex is
physically large, about 2.75 times larger than that of a chimpanzee,
but has only 1.25 times as many neurons.}  It is about the size of a
large handkerchief (about 2.5 sq.\ ft.) and about a tenth of an inch
(3mm) thick.  It comprises six layers (three in a few areas), each
with distinct types and arrangements of neurons and is physically
uniform across most of its area.  However, generalized connectivity
would lead to long-distance and presumably slow connections, so the
cortex is organized into tightly connected ``columns'' about 1/25th
inch (1mm) in diameter.  Each column has about 90 thousand neurons and
perhaps 250 million synapses \cite{Presti16}.  Adjacent columns may
respond to different stimuli (e.g., different parts of the retina) but
are generally part of an area that performs some specific function
(for example, language expression is mostly performed by Brocas' area
in the left hemisphere).

Modern neuroscience reveals amazing sub-specialization in cortical
areas.  For example, the primary visual cortex has dozens of
specialized subfunctions, some of which recognize edges at particular
orientations, while others recognize outlines.  The areas are highly
connected internally, with relatively sparse connections to other
areas \cite{Gazzaniga:who12}.  A deep question for understanding
consciousness is how do all these separate, fragmentary, perceptions
come together to create our unified experience?  The natural
speculation that there is some central ``homunculus'' where it all
comes together seems to be excluded by current understanding of
neuroanatomy.  It would then seem that the unified experience must be
created by some distributed computation, and that surely requires some
long distance communication across the distributed areas of
specialization.  This might be accomplished by neurons with long-range
axons, or by electrical waves.\footnote{Snakes do not have a unified
representation of the world: to a snake, a mouse is many different
things with no ability to transfer or fuse information from different
senses.  Striking the mouse is controlled by vision, finding its body
is controlled by smell, and swallowing it by touch.  A snake that has
a mouse in its coils will search for it as if it had no information
\cite{Sjolander99}.  It would be interesting to learn what features
are absent from snake brains.}

The search for Neural Correlates of Consciousness (NCC) attempts to
identify regions or functions of the brain that seem correlated with
these and other mechanisms or indicators of consciousness
\cite{Koch-etal16:NCC}.  There is scarcely any region, or structure, or
pattern of electrical activity in the brain that has not been
identified in NCC studies.  However, strong candidates include the
fronto-parietal cortices, high-frequency electrical activity in the
gamma range (35--80 Hz), and the occurrence of an EEG event known as
the P300 wave.  Although much has been learned about these and other
possible correlates, it yields little insight on the nature or
mechanisms of consciousness.

Psychology does provide some clues.  Much research and speculation in
psychology focuses on the functional organization of the brain above
the anatomical level.  There is no doubt that much of what the brain
does is unconscious, so one question concerns the allocation of
functions between conscious and unconscious.  The popular
``Dual-Process'' theory identifies two cognitive (sub)systems
\cite{Frankish10,Evans&Stanovich13,Kahneman11}: \emph{System 1} is
unconscious, fast, and specialized for routine tasks; \emph{System 2}
is conscious, slow, easily fatigued, and capable of deliberation and
reasoning---it is what we mean by ``thinking.''  However, much of this
thinking by System 2 subconsciously recruits the capabilities of
System 1, as when we make a ``snap decision'' or ``trust our gut
instinct.''

The ``Wason selection task'' illustrates some of these topics
\cite{Wason68,Cosmides89}.  Four cards are presented to the subject,
each has a color patch on one side and a number on the other and the
subject is asked to indicate which cards must be turned over to test
truth of the proposition that if a card has an even number on one
side, then it has a red patch on the other.  Thus, shown four cards
displaying 15, 18, red, and blue, the correct answer is the 18 and the
blue.  Less than 10\% of subjects perform this task correctly.  But if
the color patches are replaced by pictures of either beer or a soft
drink and the subject is asked which cards to turn over to validate
the rule ``you must be 18 or over to drink alcohol'' then a majority
can perform it correctly (i.e., select the card with 15 and the one
showing beer).

One interpretation of these results is that the first, abstract
version of the task requires System 2, which can reason about novel
problems, but whose capabilities are difficult to exploit without
training (e.g., in this case, logic), whereas the second, concrete
version can recruit built-in System 1 capabilities for policing social
rules.

This raises the question, what kinds of reasoning are supported by
built-in System 1 capabilities?  Evolutionary psychology posits that
just as evolution delivered specialized physical organs such as hearts
and livers, so it leads to specialized mental ``modules'' such as
those for social rules, mate selection, and so on.  An alternative
view is that System 1 capabilities are primarily concerned with our
embodiment in the physical world, and focus on distance, weight, time
etc., but other topics can be mapped onto these basic ones, which
become metaphors when expressed linguistically.  Thus, social
organization is mapped onto elevation: ``he is my superior, while she
is my peer and I am above the others.''  Lakoff and Johnson
\cite{Lakoff&Johnson08} posit that this application of metaphor
pervades our thinking: it is a (primary) mechanism of thought, not
just a feature of language.  ``Mental models,'' which underpin our
thinking about complex systems and artifacts \cite{Craik43}, can be
seen as deliberately constructed bridges between System 2 and System 1
capabilities.

A unique human attribute, and the reason we dominate the world, is our
ability to engage in collaborative activities with joint goals and
intentions; this is referred to as \emph{shared intentionality}
\cite{Tomasello-etal:intentions05}.  Social insects and animals also
behave collaboratively and cooperatively, but these behaviors are
programmed by evolution: a chimpanzee cannot create a new goal and
communicate it to its fellows, so ``it is inconceivable that you would
ever see two chimpanzees carrying a log together'' \cite[quoting
Tomasello on page 238]{Haidt13}.

Communication and cooperation require an ability to see things from
the other party's point of view: I cannot explain something to you
without giving some consideration to your existing state of knowledge
and belief.  This is known as a \emph{theory of mind} and develops in
human children at about the age of four.  A common way to examine this
is the ``false belief test.''  Here, a child watches a puppet
performance in which a puppet enters and places a toy in a box, then
exits; a second puppet enters and moves the toy to a different box,
then exits; the first puppet reenters and the child is asked where it
will look for the toy.  Young children will say the second box: they
know that is where it is and have not yet grasped that others may have
different information and beliefs.  Four year olds will identify the
first box.  This test has also been applied to animals and the results
are debated \cite[and subsequent responses]{Krupenye-etal16}.

A theory of mind requires perception of others as independent agents
with their own beliefs, desires, and intentions; this presumably
builds on, or leads to, the recognition of ourselves as just such an
agent.  This is \emph{self-awareness} (it should really be called
self-consciousness but this term is commonly used with a different
meaning) and seems to be one of the harbingers of consciousness, so we
will now move from what is known about the brain to what is postulated
about consciousness.

\section{Theories of Consciousness}
\label{theories}

There are dozens of theories of consciousness; we divide them into
three main groups, which we term \emph{universalism},
\emph{biologicism}, and
\emph{functionalism}.\footnote{\emph{Illusionsism} \cite[and other
papers in the same issue]{Frankish16} constitutes another group of
theories.  We do not consider them here because they provide little
insight on the possibility of machine consciousness (the speculation
is that machines will need to suffer the same illusions as humans, so
first we need to understand the human illusion of consciousness).}  In
this section we outline these groups and sketch some representative
theories from each.  We do not go into any detail because our purpose
is simply to indicate the wide range of current theories.

Theories that we group under the name universalism hold that
consciousness is an intrinsic feature of the universe, rather like
evolution, and may be found wherever certain conditions obtain.  The
broadest theories of this kind are various forms of
\emph{panpsychism}, which hold that everything in the universe,
including elementary particles, is conscious to some degree
\cite{SEP:Panpsychism}.  One attraction of panpsychism is that
human-level consciousness can then be seen as simply the ``sum'' of
its constituents, without the difficulty of explaining its emergence
from unconscious antecedents.  However, panpsychism smacks of dualism
so it needs to explain how universal consciousness gains purchase on
the material world.  Of course, panpsychism has other difficulties, not
the least of which is that most people view it with
incredulity.\footnote{This was also once true of plate tectonics, the
microbial cause of gastric ulcers, and the symbiotic origin of
mitochondria and chloroplasts.}

More structured theories of this kind see consciousness as a natural
consequence of complex systems whose integration makes them more than
the sum of their parts.  In other words, consciousness derives from
integration in complex systems.  We can suppose these systems have
some richness of ``behavior'' or ``knowledge'' or \ldots something;
the concept most generally used is \emph{information}.  Some of this
information will be due to the constituent parts of the system, and
some of it will be due to their integration and interaction.  We can
attempt to measure the latter by partitioning the system into some
number of disconnected parts and then taking the difference between
the information in the original system and the sum of that in its (no
longer integrated) parts.  This is the basis of \emph{Integrated
Information Theory} (IIT) \cite{Tononi-etal16:PSC} and (after taking
care of a lot of details that are glossed over in the sketch just
given) it provides a measure known as $\Phi$.

The interpretation is that high $\Phi$ corresponds to consciousness.
It is not obvious why this should be so, nor how the abstract model
should apply to humans (and there are several other objections
\cite{Cerullo15:problemPHI}) but Giulio Tononi, the primary author of
the theory, takes pains to explain how it is based on phenomenological
considerations (abstracted as five axioms)
\cite{Tononi&Koch18:everywhere} and can illuminate the search for a
physical substrate of consciousness in the human brain
\cite{Tononi-etal16:PSC}.  A physical test for assessment of
consciousness, inspired by IIT, reliably discriminates the level of
consciousness in individuals during wakefulness, sleep, and
anesthesia, as well as in patients who have emerged from coma
\cite{Tononi13:PCI}.  Recent studies indicate that $\Phi$ also can be
used to predict performance in group activities: for example, high
$\Phi$ among Wikipedia editors is correlated with higher quality
articles \cite{Engel&Malone18}.

Since universalist theories hold that consciousness is everywhere, it
might seem that they are supportive of machine consciousness, but this
does not seem to be so---because, with the exception of IIT, they do
not explain how primitive consciousness is aggregated so that the
human brain has a lot of it, whereas a comparable mass of straw does
not.  The basic formulation of IIT does seem to allow machine
consciousness, but also suggests that the United States of America
should be a conscious entity, so recent treatments have some
adjustments.  These favor feedback networks (as found in the brain)
and discount feed-forward; technological systems, whether based on
ANNs or conventional computation are asserted to be equivalent to
feed-forward networks and machine consciousness is thereby denied
\cite{Tononi&Koch18:everywhere}.

While universalism finds consciousness among properties of the world
at large, biologicism locates it in the biology of the brain.  A
theory known as \emph{Orchestrated Objective Reduction} (Orch-OR)
combines the two classes of theories \cite{Hameroff&Penrose13}.
Orch-OR asserts that the human mind, and consciousness in particular,
cannot be explained by functional or computational models of the
brain.\footnote{This argument is developed by Penrose \cite{Penrose94}
using a contentious interpretation of G\"{o}del's first incompleteness
theorem; Feferman provides a rebuttal \cite{Feferman95:Penrose} in a
journal issue devoted to critical commentaries on Penrose' argument.}
Therefore some extra ``spark'' is required and it is proposed that
this comes from collapse of the quantum mechanical wave function,
which must take place at exceedingly small physical
dimensions.\footnote{Several of the founders of quantum mechanics
explicitly discussed its possible relationship with
consciousness \cite{Marin09}.}  Stuart Hameroff proposed that the
microtubules within neurons can perform this quantum processing and
thereby provide noncomputational decision making or awareness that
interacts with conventional neural activity to produce
consciousness \cite{Hameroff&Penrose13}.

While Orch-OR combines universalism and biologicism, \emph{Global
Workspace Theory} (GWT) combines elements of biologicism and
functionalism \cite{Baars05:GWT}.  The functionalist aspect describes
an architecture for mental activities similar to the blackboard
architecture of AI: many unconscious mental processes read and write to
a global working memory that is selectively attended to; consciousness
corresponds to a ``spotlight'' of attention on this global workspace.
The biological aspect associates various elements of brain physiology
(e.g., cortical areas, gamma synchrony) in the realization of this
architecture.

Biological theories have impact on the possibility of machine
consciousness to the extent that their biological content is essential
rather than incidental.  Since nothing in a computer resembles
neuronal microtubules, machine consciousness is ruled out by the
biological aspect of Orch-OR.  But that may not eliminate artificial
consciousness altogether since we might devise some alternative
structure to collapse the wave function, though it would not emerge
from conventional computation.  Dually, if its cognitive architecture
is the key element of GWT, then reproducing this architecture in
computation could enable machine consciousness, but this might not be
so if some aspects of the biological implementation are critical.

Functionalist theories deny any special role for biology and assert
that consciousness arises from what the brain does.  Unsurprisingly
(since nothing was known about the biology of the brain until
recently) functionalist theories of consciousness are among the
oldest, dating back to William of Ockham or even Saint Augustine
\cite{Brower-Toland12}.  The prototypical functionalist theories are
various forms of \emph{Higher Order Thought} (HOT)
\cite{Gennaro04,Rosenthal04}, that is, thoughts about thoughts:
``consciousness is the perception of what passes in a man's own mind''
(John Locke).

In the previous section, we noted that humans have self awareness;
related to this is introspection, which is the mind observing its own
(conscious) operation and thus an instance of higher-order thought.
In fact, this seems to be a second level of higher-order thought: the
conscious mind is thinking about its own conscious operation.  A first
level of higher-order thought would seem to be basic awareness: a
conscious thought about something presented by the unconscious.  For
example, I am conscious of the coffee cup on my desk; the sight of the
coffee cup, along with many other items on my desk, is presented by my
unconscious visual system, and the coffee cup is the one of which I am
conscious: so I have a (first level) higher-order thought directed at
the (base level) mental representation of the cup.  Some classical HOT
theories posit that a thought becomes conscious when it is the target
of a higher-order thought.  This leads to an objection that thinking
of a rock surely does not make the rock conscious and there is a lot
of theoretical wriggling to deal with this \cite[Section
4.3]{Gennaro2012}.  As computer scientists, we would assume it is the
higher-order thought that is conscious, not the base-level target, so
this ``problem of the rock'' seems specious.

Computer science provides some other ideas and terminology that can be
useful here.  Much of the function of the brain is to provide a
control system for the body: given a goal and sense input, calculate
actions to accomplish the goal.  The goal may be to catch a fly ball,
cross the street, win an election, or survive until your offspring are
independent.  Computer control systems do similar (though generally
simpler) things and have a certain structure.  In particular, they are
always based on a model of the controlled ``plant'' and its
environment \cite{Conant&Ashby70}.\footnote{Conant and Ashby
explicitly recognized this must apply to the brain, which seems
remarkably prescient for 1970: ``The theorem has the interesting
corollary that the living brain, so far as it is to be successful and
efficient as a regulator for survival, must proceed, in learning, by
the formation of a model (or models) of its environment.''}  For
elementary control systems, such as the cruise control in a car (or
humans catching fly balls), the model is used in design (or implicitly
``discovered'' during evolution) of the system but is not explicitly
represented in the implementation.

More complex control systems do have an explicit model represented in
the deployed system and may (in adaptive systems) have ways to adjust
its parameters during operation.  Even more ambitious systems may
learn and adapt the model as they go along.  \emph{Reflective} systems
are those that construct an explicit model of their own operation and
use this to adjust future behavior.  When computer scientists are
asked what consciousness might be, they are very likely to answer that
it must be something like reflection, and most efforts to construct
machine consciousness are based on reflection.

\emph{Predictive Processing} (PP) \cite{Clark13} is a recent, and
increasingly popular, theory of brain operation that posits that every
level and every subfunction of the brain has a model and uses (an
approximation to) Bayesian estimation to minimize prediction error, as
was described for sensory interpretation in the previous section.  In
total, the brain learns and maintains models for various aspects of
the world and uses sense input and experience to make these as
accurate as possible.  PP sees consciousness less as a ``thing'' and
more as the continuous process of building and updating models.
(``Free Energy'' \cite{Friston:free-energy10} is a more
all-encompassing variant that includes actions: the brain ``predicts''
that the hand, say, is in a certain place and to minimize prediction
error the hand actually moves there.)

Predictive processing could enable successful individual behavior
without a separate mechanism for consciousness.  But, as noted
earlier, a distinguishing feature of humans is the creation of shared
intentionality.  If you and I are to cooperate on a common goal, then
we need to share similar models for some aspects of the world---and
one interpretation of consciousness is that its function is to
construct higher-order thoughts that are succinct descriptions or
explanations of unconscious models suitable for communication to
others.

This rapid and superficial overview of theories of consciousness
reveals a disappointing truth: there is a vast number of theories and
for the most part they are mutually incompatible.  It follows that
most of them must be wrong, although few are presented with the
specificity to support falsifiable
experiments.\footnote{Falsifiability is the touchstone that separates
scientific theories from speculation.  This criterion is due to Popper
\cite{Popper:2014logic} and is generally endorsed by working
scientists (as a declaration of what science \emph{is}, not
necessarily how it is \emph{done}).  However, it is hotly debated,
even rejected, by philosophers of science: ``if Popper is on the right
track, then the majority of professional philosophers the world over
have wasted or are wasting their intellectual careers''
\cite{Bartley76}.  Most theories of consciousness do not propose
falsifiable experiments and are therefore, in the words of Wolfgang
Pauli, ``not even wrong.''}$^{,}$\footnote{\textbf{Update:} Yaron and
colleagues recently developed a database that culls results from 379
papers reporting 418 experiments\cite{Yaron-etal:database21}; they
conclude that ``the field generally suffers from a strong
confirmatory-bias, and that the majority of studies post-hoc interpret
their findings concerning the theories, rather than designed a-priori
to test their critical predictions'' \cite{Yaron-etal21}.  New
activities, however, do report indications of possible falsifications:
for Orch-OR
\url{https://physicsworld.com/a/quantum-theory-of-consciousness-put-in-doubt-by-underground-experiment/},
and GWT
\url{https://bigthink.com/neuropsych/revision-leading-theory-consciousness/},
respectively.  Furthermore the Templeton Foundation is funding an
effort to ``test competing predictions'' by GWT and IIT
\url{https://www.templetonworldcharity.org/projects-database/accelerating-research-consciousness-adversarial-collaboration-test-contradictory}.}  Furthermore, few (if any) theories
suggest what consciousness might be \emph{for}, or what it
\emph{does}.  Manzotti and Chella identify a yet more basic criticism,
which they call ``the intermediate level fallacy''
\cite{Manzotti&Chella18}: theories of consciousness identify various
functional or biological mechanisms (the intermediate level) that are
asserted to produce consciousness, but they fail to explain how the
mechanism produces the experience.

These criticisms do not imply that theories of consciousness are
without merit (though it is disappointing that few of them shed any
light on the counterintuitive aspects of consciousness, such as those
described in Section \ref{mystery}).  The field is so young and
opportunities for observation and experiment so few that
falsifiability may be a premature expectation.  The theories can be
seen as pointers to directions that may be worthy of further
exploration.  In the following section, we describe explorations based
on computer simulations.

\section{Attempts to Create Machine Consciousness}
\label{mc}

We have seen that there is a large number of theories of consciousness
and rather little evidence to help choose among them.  One approach is
to construct simulations; at the very least, this forces elaboration of
sufficient detail  that we can construct explicit models of the
mechanisms of the chosen theory and explore their properties.

Early experiments were conducted by Tiham\'{e}r Nemes in the 1960s
\cite{Nemes70}, but intelligence and consciousness were not sharply
distinguished in those days, nor were cybernetics and (what became)
AI\@.  In 1989, Leonard Angel published a book with the provocative
title ``How to Build a Conscious Machine'' \cite{Angel89} in which he
proposed a kind of agent system.  A modern view of machine or robot
consciousness is attributed to Igor Aleksander in 1992
\cite{Aleksander92}, who postulated that such a robot would need
representations for depiction, imagination, attention, planning, and
emotion and that consciousness could emerge from their interaction.

The first large project to explore machine consciousness was {\sc
cronus} \cite{Marques-etal07}.  This was predicated on the idea that
reflection---internal models of the system's own operation---play an
important part in consciousness.  Physically, {\sc cronus} was an
\emph{anthropomimetic} robot (i.e., one closely based on the human
musculoskeletal system) with a soft-realtime physics-based simulation
of the robot in its environment.  The internal simulation allowed the
robot to project the effects of possible future actions, which the
authors describe as ``functional imagination'' \cite{Marques-etal08}.
Later studies used a yet more complex robot (``{\sc eccerobot}''),
while earlier ones used a very simple, nonanthropomorphic device
\cite{Holland03}.  It is not clear to us that complex robots added a
great deal to these experiments, and they certainly increased the
engineering challenges.

Experiments by Chella and colleagues explored robots' interaction with
others; this requires a theory of mind, and a sense of self
\cite{Chella&Manzotti09,Chella-etal08}.  Gamez describes other
projects performed around the same time \cite{Gamez08}.  All these
experiments and those mentioned above employ some form of reflection
or HOT as their underlying theory of consciousness.  Others have built
systems based on GWT or IIT; Reggia provides a survey \cite{Reggia13}.

Those who subscribe to biologicist theories of consciousness might
argue that functionalist approaches are doomed to failure and that
what is needed are simulations of the biology of the brain.  ``Systems
biology'' supports this by building computational (``in
silico'') models of biological processes, including neurons.
Typically, these are Matlab-type models with differential equations
for the flow of ions and various other elements in the cell.  Such
models often require experimentally-derived parameters such as rate
constants and are limited to systems involving only a few cells.
\emph{Symbolic} Systems Biology abstracts from much of this detail and
uses state machines or rewriting.

It is clear that simulations of the whole, or even significant parts
of, the human brain are computationally infeasible at the level of
detail pursued in systems biology.\footnote{The state of the art is a
few neurons; one example considers the central pattern generator (for
rhythmic gut movements during feeding) of Aplysia (a marine mollusk),
which is comprised of 10 neurons \cite{Tiwari&Talcott08}.}
Furthermore, such simulations would require detailed information on
the neural architecture of the brain that is currently lacking,
although research to develop this information is being pursued by the
Human Connectome Project.\footnote{This could be disrupted if glial
cells and their separate connectivity turn out to be significant.}  It
is possible that abstractions of biological functions of the brain can
be developed that will permit larger scale simulations
($\!\!$\cite{Grossberg17} may be a step in this direction).

Research on machine consciousness seems not to have a central forum
for presentation of results and discussion of ideas: the
\emph{International Journal of Machine Consciousness}
began publication in 2009 but ceased in 2014.  Perhaps as
a result, recent work seems to retread familiar ground.  For example,
a paper by Dehaene, Lau and Kouider from 2017 \cite{Dehaene-etal17}
presents the authors' theory of consciousness (global availability as
in GWT, plus reflection built on PP), then asserts that a machine with
these capabilities ``would behave as though it were conscious''
\cite{Dehaene-etal17}.  In a response, Carter \emph{et al.}\
\cite{Carter-etal18} observe that Dehaene and colleagues ask and
answer the wrong questions---essentially, Dehaene \emph{et al.}\ are
aiming for intentional consciousness, whereas Carter \emph{et al.}\
think that phenomenal consciousness is what matters: for machines to
be conscious, ``we must ask whether they have subjective experiences:
do machines consciously perceive and sense colors, sounds, and
smells?''  They posit ``a more pertinent question for the field might
be: what would constitute successful demonstration of artificial
consciousness?''

These are old questions (e.g., \cite{Boltuc09}) and it is
disappointing that the dialog does not seem to be moving forward.  A
more basic question, given that researchers do not yet claim to have
demonstrated machine consciousness, asks what should we expect to
learn from research on machine consciousness, whether based on
functionalist or biologicist principles?  As with AI, researchers
distinguish \emph{strong} and \emph{weak} forms of machine
consciousness (sometimes framed as duplication vs.\ simulation).
Strong machine consciousness would \emph{be} conscious, whereas the
weak form is a philosophical zombie: it exhibits behaviors and
attributes associated with consciousness without actually possessing
it.  We believe it is fair to say that existing attempts to create
machine consciousness have not expected to achieve the strong form.
Rather, they simulate selected mechanisms from a theory of
consciousness and observe resultant behaviors and properties.  What
these studies of weak machine consciousness can achieve is to frame
and explore precise statements of the form: ``in a system of type W,
some components X will manifest characteristics of type Y under some
circumstances Z'' (Owen Holland).

There is no doubt that experiments of this kind can help sharpen and
discriminate among various theories of consciousness, but most
observers are skeptical that the weak form of machine consciousness
leads to the strong.  By analogy, we can build extremely accurate
simulations of the cosmos and explore the missing mass (due to dark
matter and dark energy), yet the simulations do not \emph{have} mass;
so a simulation \emph{of} consciousness will not \emph{have}
consciousness.  On the other hand, the weak and strong distinction
seems really to matter only for phenomenal consciousness and its
cohorts: we likely will regard an entity that \emph{has} feelings
differently than one that merely simulates them.  But weak intentional
consciousness is operationally equivalent to the strong: if it enables
some new cognitive capabilities then the underlying technology can use
these by running the weak simulation as a subroutine.  This asymmetry
between phenomenal and intentional consciousness is related to the
``hard problem'' of consciousness \cite{Chalmers95} and may be a
crisper way to formulate it than the traditional description.

The topic that motivated this study is concern that we may
(inadvertently) develop technology with strong machine consciousness.
We explore that possibility in the following section.

\section{Possible Emergence of Technological Consciousness}

In this section we explore the possibility that consciousness might
emerge from a technological substrate rather as natural consciousness
emerges from the brain, but first we should examine this notion of
``emergence'' in a little more detail.

Philosophers ascribe different interpretations to this notion, and the
interpretations have changed over time.\footnote{There was a whole
school of ``British Emergentists'' in the late 19th century.}  The one
that is most germane to our discussion is generally called \emph{weak}
emergence \cite{Bedau97}.  In all discussion of emergence, there are
``upper level'' \emph{macro} phenomena that arise out of ``lower
level'' \emph{micro} phenomena.  What characterizes emergence as
opposed to mere assembly is that the macro phenomena are of a
\emph{different kind} than the micro phenomena and are described with
a different vocabulary than the micro: that is, they are
``ontologically novel'' and, if understood in sufficient detail, are
characterized by different models and mathematics.  A classic example
is pressure and temperature in a gas (macro), which emerge from the
motion of its molecules (micro).  Macro level phenomena that are not
ontologically novel are called \emph{resultant}.  Consciousness arises
from the brain yet is ontologically distinct, so it is legitimate to
speak of it emerging from the brain.

\emph{Weak} emergence is characterized by the requirement or
expectation that macro phenomena can be \emph{explained} in terms of
micro.  We may be able to do this precisely (by ``bridge laws'') as in
the case of gas properties, or we may lack the knowledge to do so, as
in the case of consciousness emerging from the brain, but we are
confident such an explanation exists.  This contrasts with
\emph{strong} emergence, which asserts that macro phenomena cannot be
(completely) explained in terms of micro.  Weak emergence is
compatible with materialism, while strong emergence is tantamount to
property dualism.

There are further concepts that can be added to a theory of emergence,
such as \emph{supervenience} (if two macro states differ, they must
arise from different micro states---i.e., there is a mathematical
\emph{function} from micro to macro), and some of them, such as
\emph{downward causation} (where a change in macro states causes a
change in micro states, contrary to the notion that macro states arise
out of micro), are philosophically challenging
\cite{Campbell&Bickhard11}, but the basic ideas and terminology are
sufficient for our purposes.

In this section we are concerned with the possibility that
consciousness might emerge ``accidentally'' from the ever more complex
technology that we construct, especially as some of it is explicitly
inspired by human mental capabilities and by approximations to some of
its mechanisms, such as Artificial General Intelligence, deep
learning, and so on.

Although the small-scale experiments in machine consciousness outlined
in the previous section seem not to have produced (even weak)
consciousness, the sheer quantity of computational power that is now,
or soon will be, available, might raise concern.  In Section
\ref{brain} we sketched the amazing scale and complexity of the human
brain and noted that it has about $10^{14}$ synapses, which we can use
as a proxy for its ``capacity.''  It is difficult to say how many
transistors or computational gates or memory bits are equivalent to a
synapse (temporarily adopting a computationalist theory of
consciousness), but we can observe that the available number of these
technological elements has increased exponentially---roughly ten-fold
every four years---since 1966, so that by 2026 it will have increased
by $10^{15}$ over that period, and this factor dominates any starting
assumption.  It seems plausible, therefore, that technology will have
raw computing capacity roughly equivalent to the human brain by about
2026 (if not then, just wait another four
years).\footnote{\textbf{Update:} Large language models (essentially
neural nets trained to predict the next word in a stream of text) have
increased in scale at an astonishing rate.  In 2018, the first
Generative Pre-Trained Transformer (GPT-1) from OpenAI had 117 million
parameters, GPT-2 a year later had 1.5 billion, and GPT-3 in 2021 has
175 billion.  LaMDA, the model that Blake Lemoine declared
``sentient'' has 137 billion parameters, and Google and others are
said to be working on models with more than a trillion parameters,
which is believed to approach the scale of the human brain.  Some of
the largest supercomputers in the world are used for training these
and other neural nets (e.g., for self-driving cars).  The performance
of these systems appears to have grown with their scale and most
people are astonished at the quality of the text they can generate.
This admiration is not universal, however, especially among those who
``know how the trick is done'' and who dismiss these systems as
``stochastic parrots''
\url{https://medium.com/@emilymenonbender/on-nyt-magazine-on-ai-resist-the-urge-to-be-impressed-3d92fd9a0edd}.}
Calculations such as this are the basis for speculation on ``The
Singularity'' \cite{Kurzweil10}; the important point is that under
almost any assumptions, exponential growth guarantees technological
computational capability will match that of a human brain within a
decade or so.

On the other hand, none of the theories of consciousness surveyed in
Section \ref{theories} gives reason to suppose that scale makes
consciousness more likely: they each identify consciousness with some
mechanism or structure that is either present or absent.  (IIT is
sensitive to scaling but excludes machine consciousness on other
grounds.)  We have to conclude, therefore, that current experiments
and existing theories of consciousness do not provide much guidance on
the possible emergence of strong consciousness in large scale
technological systems.

The lack of commonality among the theories means that even if we
ignore their details and look for broad indications, they still
provide few clues where to look for possible emergence of machine
consciousness.  If forced to nominate one idea that informs all or
most of the various theories, we would choose ``integration.''  This
is explicit in IIT, implicit in GWT and other biologicist theories,
and explicit again in some HOT theories---although it is not clear
they agree on what it is that is integrated.  The archaeologist Steven
Mithen attributes the ``explosion'' of creativity seen in the human
record about 50,000 years ago (cave paintings etc.)\ to integration of
formerly separate cognitive domains \cite{Mithen96}.  This idea is
compatible with the metaphor theory: we became able to think
effectively about many new topics once we acquired the ability to map
them to (previously separate) built-in capabilities.

If integration of separate cognitive domains is a pointer to human
consciousness, might something similar occur in large scale
technological systems?  At present, our technical systems tend each to
focus on a narrow range of functions, defined by their task and
purpose.  A self-driving car just drives: it does not concern itself
with raising offspring, finding a mate, or maintaining status; nor
does it vacuum the floor, wash clothes, or guide airplanes.  It is
possible that as the Internet of Things develops, the self-driving car
will be able to integrate, on some level, with robot vacuum cleaners
and washing machines, and with the air traffic control system.  In a
foreseeable future, it may integrate with babysitting ``carebots'' and
dating sites, and may also have a reputation index that it is
programmed to defend.  The total system will then have many different
skills and capabilities and their integration may yield something new.
We cannot predict whether this might include any form of
consciousness\footnote{Discussion around Searle's ``Chinese Room''
might be relevant here \cite{Chinese-room02}.}  but it seems possible
that it could lead to coordinated behavior that resembles that
associated with consciousness.\footnote{The skeptic might think it is
more likely to lead to ``emergent misbehavior'' \cite{Mogul06}.}

In May 2001, The Swarz Foundation sponsored a workshop on the topic
``Can a Machine be Conscious?'' at Cold Spring
Harbor.\footnote{\url{http://www.theswartzfoundation.org/banbury_e.asp}}
The attendees concluded ``we know of no fundamental law or principle
operating in this universe that forbids the existence of subjective
feelings in artifacts designed or evolved by humans.''  We would
agree, though generalized from ``subjective feelings'' (i.e.,
phenomenal consciousness) to include intentional consciousness as
well.  But just as we know no reason that excludes technological
consciousness, the discussion above suggests we also have no reason to
think that its emergence, at least in a strong form, is imminent or
likely.

Since we cannot know whether a technological consciousness might
emerge, nor what form it might take, we consider a range of
possibilities.  Gamez \cite{Gamez08} identified four different topics
within research on machine consciousness.  We adapt his topics to our
discussion.

\subsubsection*{TC1: Technology with external behavior associated with
consciousness}

This is essentially the goal of Strong AI, or AGI\@.  It presumes that
consciousness is inessential to the construction of (some or all)
human-level behavior: consciousness may be the way humans achieve
certain capabilities, but technology might achieve them by other means.

This, or an approximation, seems a plausible possibility, and raises
well-known concerns about the hazards and safety of advanced AI
\cite{Everitt-etal18}.  The ``integration'' scenario outlined above
adds another twist that could raise concern further.

\subsubsection*{TC2: Technology with cognitive characteristics
associated with consciousness}

Exactly which cognitive characteristics are of greatest interest is a
topic for debate, but candidates include the ability to construct
shared intentionality (i.e., engage in teamwork), or to have original
ideas.  Much of the concern about AGI centers on the idea that it
could be capable of self-improvement, and would thereby outstrip human
capabilities at an increasing pace.

Again, it seems plausible that these capabilities can be constructed
by technological means, without consciousness.  Recent work on
automation of scientific discovery \cite{Schmidt&Lipson09} suggests
that brute search over internal models can accomplish some of what was
thought to require imagination, insight, or originality.

\subsubsection*{TC3: Technology with an architecture that is
    claimed to be a cause or correlate of human consciousness}

Whereas technology constructed for other purposes, no matter how large
or complex, might not achieve consciousness, it is possible that
technology that reproduces or simulates salient architectural or
biological aspects of the human brain might do so.

This possibility does not seem imminent, in that we really have no
idea what the salient aspects of the human brain might be; indeed the
purpose of much of the research on machine consciousness discussed in
Section \ref{mc} is to use computer models and simulation to help
refine and evaluate (and possibly refute) proposed theories of
consciousness.  Rather than harbingers of technological consciousness,
this research could be seen to provide reassurance that some
architectures that might raise concern, such as reflective systems
with sophisticated self models, have not exhibited consciousness.

Furthermore, as noted in Section \ref{mc}, while it is possible that
TC3 systems might eventually simulate consciousness, this does not
imply that they will \emph{be} conscious---that is the concern of TC4.

\subsubsection*{TC4: Technology that truly is conscious}

The topic of concern here is strong machine consciousness.  This is
relevant only for phenomenal consciousness because, as explained in
the previous section, weak and strong are equivalent for intentional
consciousness---and furthermore, as noted for TC1 and TC2, it is
plausible that advanced AI can be achieved without any kind of
technological consciousness.

Most observers believe that strongly phenomenally conscious technology
is possible.  (There are dissenters, most of whom deny
computationalism \cite{Bringsjord&Zenzen97}, or endorse Ross' claim
for the immateriality of thought \cite{Ross92}.)  However, the
experiments in machine consciousness outlined in the previous section
do not suggest that it is imminent.  But on the other hand,
technological consciousness might be nothing like human consciousness
and could emerge from entirely different mechanisms.  It seems that at
present we must accept that we just do not know enough to evaluate
this concern.

We have considered a range of capabilities that might emerge; we now
examine how we might detect and measure it.

\section{Detection and Measurement of Consciousness}
\label{measurement}

Detection of consciousness is difficult because a) we do not know what
consciousness \emph{does} nor what it is \emph{for}, and b) we do not
know how it comes about.  Thus, we do not know what functions or
behaviors or capabilities to look for, nor what mechanisms or
precursor functions.  Hence, most proposed tests for consciousness
amount to little more than attempts to observe human-like capabilities
that are thought to require or to be associated with
consciousness.\footnote{\textbf{Update:} Large language models such as
GPT-3 and LaMDA already come very close to passing the ``Turing Test''
where an interlocutor is unable to distinguish a dialog with a machine
from one with a human.  This suggests that apparent facility with
language may not be an effective test for consciousness.}  Standard
examples include various embellishments of the Turing Test, such as
the Total Turing Test (TTT) \cite{French00} and variants specialized
toward phenomenal consciousness (e.g., wine tasting).  Many of them
assume the ability to converse in natural language, or use that as a
diagnostic.

Although intended to be applied to machines, tests that seek adult
human-like capabilities will automatically conclude that animals and
human infants lack consciousness.  Yet many believe that their
consciousness is at least a possibility.  We refer to these tests as
\emph{anthropocentric} and consider them unlikely to detect a
machine consciousness that emerges unexpectedly and may have little
resemblance to our own.

On the other hand, given that adult humans are the only subjects that
we know to possess consciousness, it is difficult to propose neutral
tests.  One illustrative example is whether a subject that has learned
a skill will teach it to others.  A chimpanzee that has learned to use
a stick as a tool to fish for ants does not attempt to teach others:
it seems unconscious (unaware) that it has this skill.  Since
chimpanzees are social animals, this seems like a legitimate test.
However, it would not seem legitimate for solitary creatures, such as
octopuses.  And what would it reveal about technology?

Raoult and Yampolskiy \cite{Raoult&Yampolskiy15} perform a thorough
literature survey and identify 21 different tests aimed at machine
consciousness.  Most are anthropocentric; a few, such as variants on
the Mirror Test (does the subject recognize itself in a mirror?) are
debatable, while others, such as attempts to measure $\Phi$, seem
truly neutral.

There is obvious tension between tests that are too hard and too easy.
Excessively anthropocentric tests will fail to detect alternative
forms of consciousness, so may be considered too hard.  On the other
hand, the Mirror Test seems too easy as the capability is readily
programmed in computers \cite{Bringsjord-eta15}.  Another test that
seems too easy, despite its apparent sophistication, is Insight
Learning (the ability to solve problems by making new relations
between previously acquired knowledge rather than through trial and
error): this skill is easily programmed (search over internal
simulations) with no requirement for consciousness.

A related objection to many of these tests is that they will not
distinguish strong machine consciousness from weak: that is, they will
not distinguish a technology that truly \emph{is} conscious from a
philosophical zombie that merely simulates it.  One, rather
controversial, idea that might do so is to observe whether the system
shows interest in exploring or manipulating its own ``consciousness'':
this rests on the observation that all human societies investigate
altered states of consciousness through chemical and spiritual means.

If consciousness is hard to detect, it is even harder to measure.
Raoult and Yampolskiy identify four tests that produce numerical
scores; one of these is $\Phi$, the others are strongly
anthropocentric.  Although $\Phi$ may measure some aspects of
consciousness, it does not seem specifically sensitive to phenomenal
consciousness, and it discounts the claimed feed-forward nature of
conventional computation \cite{Koch&Tononi17}.

We conclude there are no good tests for detecting or measuring
phenomenal consciousness, especially incarnations different to our
own.  Detection and measurement of intentional consciousness may be
unnecessary: what we care about are the capabilities and behaviors
that it enables, and these will be observed directly.

\section{Ethics for Control of Conscious Technology}
\label{control}

A conscious technology might have goals and priorities that conflict
with those of human society.  So it seems prudent that technology on a
path that could lead to consciousness should have overarching
constraints built in from the very beginning to forestall this danger.
We cannot know the particular circumstances that may arise, so the
constraints need to be general and overarching, rather like Asimov's
``Three Laws of Robotics.''\footnote{1: A robot may not injure a human
being or, through inaction, allow a human being to come to harm; 2: A
robot must obey orders given to it by human beings except where such
orders would conflict with the First Law; 3: A robot must protect its
own existence, as long as such protection does not conflict with the
First or Second Law \cite[The Three Laws appear in the story
``Runaround'']{Asimov50}.}  Asimov's laws were a plot device and his
stories often concern unintended consequences of these
plausible-sounding laws, thereby indicating that construction of
suitable constraints may be challenging.

One idea is that the constraints should be based on human ethics
\cite{Yu-etal18}; a dissenting opinion, advocating explicit reasoning
about safety is provided by Yampolskiy \cite{Yampolskiy13}.  Of
course, ethics have been studied and debated for millennia, without
achieving consensus.  Nonetheless, some broad general principles are
known.  Ethics are the basic rules by which societies maintain order
and cohesion; however, some very successful societies have elements
that others find repugnant: for example, Ancient Greece used slaves
(Aristotle wrote of ``natural slaves'') and Ancient Rome had execution
as a form of public entertainment.  Hence, it seems that the moral
foundations of ethics are not universal.  Nonetheless, modern
``experimental ethics'' finds that human moral sense is built on five
basic principles and those do seem universal: care, fairness,
loyalty/ingroup, authority/respect, and sanctity/purity
\cite{Haidt13}.  What is not universal is preference and weighting
among the principles, which behave rather like the five basic senses
of taste: different societies and individuals prefer some, and some
combinations, to others.  For example, western liberals stress
fairness while conservatives favor authority.

Even if an agreed weighting of the basic principles were built in to
advanced technology, it may not be obvious how to apply it.  For
example, a self driving car might be confronted by a vehicle crossing
against the lights and the choices are to crash into it, likely
killing or injuring the occupants of both vehicles, or to swerve onto
the sidewalk, likely killing pedestrians.  The fairness principle
might argue that all lives are equal and utilitarianism might then
suggest a decision that minimizes the probable injuries.  On the other
hand, the care principle might argue that the system has a special
responsibility for its own passengers and should seek a solution that
minimizes their harm.

``Trolley problems'' are thought experiments used to probe human
judgments on these ethical dilemmas.  The classic problem posits a
runaway street car or trolley that is heading toward a group of five
people.  You are standing by a switch or point and can throw this to
redirect the trolley to a different track where it will hit just one
person.  Most subjects say it is permissible, indeed preferable, to
throw the switch, even though it will injure an innocent who would
otherwise be unharmed.  However, a variant on the original trolley
problem has you and another person standing by the track and suggests
that you bring the trolley to a halt, and save the five, by pushing
the other person onto the track in front of the trolley.  Most
subjects will say this is ethically unacceptable, even though it is
equivalent to the first case by utilitarian accounting.  These
examples illustrate the ``Doctrine of Double Effect'' (DDE), which
holds that it is ethically acceptable to cause harm as an unintended
(even if predictable) side effect of a (larger) good: the first case
satisfies the doctrine, but the second violates the ``unintended''
condition.

Experimental systems have been developed that can represent and reason
about ethical principles such as DDE and these have been applied to
trolley problems, including some that involve self-harm (e.g.,
throwing yourself in front of the trolley) and thereby violate the
``unintended'' aspect of DDE
\cite{Bringsjord-etal06,Govindara-etal17}.  It is claimed that fairly
sophisticated logical treatments (e.g., intensional logics,
counterfactuals, deontic modalities) are needed to represent ethical
scenarios, and these might be additional to what is needed for the
primary functions of the system (hence, must be introduced
explicitly).  Other recent work formalizes Kant's categorical
imperative (humans must be treated as ends, not as means), which
requires a treatment of causality \cite{Lindner&Bentzen18}, while
another speculates on application of ethics to autonomous cars
\cite{Kulicki-etal18}.

There is more to ethical systems than application of ethical rules:
the underlying model of the world should have a certain neutrality
that may be hard to ensure.  For example, a system that interacts with
humans may need models of race and gender.  Whether these are
programmed or learned, they may unwittingly incorporate bias.  And in
order to interact effectively, a theory of mind may need explicitly to
construct and consider biased models.  So how can we ensure that
possibly biased models do not affect outcomes?  One possibility is
that certain judgments should be invariant under different assumptions
about self and others: that is, the system should explicitly repeat
its calculations under different assumptions as a computational
approximation to Rawls' ``Veil of Ignorance'' \cite{Rawls71}.

So far (and in the workshops), we  considered only harm to
individuals.  A truly rampant technological system could pose many
other hazards: it could undermine our institutions, or our trust in
these.  Similarly, we have mentioned only logical or rule-based
representations for ethics, whereas game theory provides another
perspective.

In addition to ethics, technological systems should also follow the
laws of their community.  There is a long history of work on
formalizing and reasoning about legal systems \cite{Gardner87}.  There
will surely be circumstances where the law conflicts with some
interpretation of ethics, or with the mission objective, so a system
constrained by several such ``overarching'' frameworks must have a
means of resolving conflicts.  Individually and in total, these are
challenging objectives.

Humans, endowed with an understanding of local ethics and of the law,
sometimes make bad judgments, or resolve conflicts among competing
ethical principles, in ways that society finds unsatisfactory.
Various forms of censure and punishment provide means to correct such
errant behavior and it seems that technological systems should also be
subject to adjustment and tuning in similar ways.  An important
question then is what is the ``accounting method'' that guides such
adjustments: is it just some internal measure, or is there some
societal score-keeping that has wider significance?  In a work
commissioned by the US government during WWII, the anthropologist Ruth
Benedict proposed a distinction between ``guilt cultures'' (e.g., the
USA) and ``shame cultures'' (e.g., Japan) \cite{Benedict46}.  This
distinction is widely criticized today, but modern ``reputation
systems,'' as employed for EBay sellers, Uber drivers, and so on
(China's Social Credit system \cite{Kobie19} extends this to the whole
society), can be seen as mechanizing some aspects of shame culture and
could provide a framework for societal control of technological
systems: the idea being that the technological system should value its
reputation and adjust its behavior to maximize this.

Being held responsible for our actions and subject to punishment and
reward seems to require that we are free to act thus or otherwise.  We
generally assume that humans have free will, but what about
technological systems?  And if they do not have free will, can they be
subject to the constraints of ethics?

Human free will is an immensely difficult subject (the most
contentious problem in all of metaphysics, according to Hume).  It is
next to impossible to reconcile the commonsense (``libertarian'' or
``contra-causal'') notion of free will---that our decisions are
uncaused causes---with materialism: in a material universe, what
happens next is determined (possibly only probabilistically due to
quantum effects)\footnote{Contrary to some na\"{\i}ve claims, quantum
randomness or other sources of nondeterminism \cite{Greenblatt:swerve}
do not open the door to free will: a random or nondeterministic choice
is no more free than a deterministic one.}  by what happened before,
so how can that determinism be suspended while I make a decision?  I
am part of the material world, so my decision is determined by my
current state (or is subject to quantum randomness) and it only
``feels like'' I made a free choice: ``all theory is against the
freedom of the will; all experience for it'' (Dr.\ Johnson).
Consequently, most philosophers accept only a weaker form of free will
(``compatibilism'') in which our conscious decisions do cause our
actions, but those decisions are not themselves uncaused causes.  In
other words, nothing prevents our deciding on one thing or the other,
but the actual choice is (probabilistically) determined: ``we can do
what we will, but we cannot will what we will'' (Schopenhauer).

Nonetheless, in everyday life we still attribute human acts to free
will and we praise or punish accordingly.  Philosophers accept this as
a useful fiction, for experiments show that subjects primed to see
free will as illusory are more likely to misbehave \cite{Cave16}, or
to become fatalistic.  Now, if it is hard to impute free will to
humans, it is even harder to impute it to technology (for what is it
but a pile of algorithms?).  However, as with humans, it is a useful
fiction to treat conscious technology (or any complex system endowed
with learning) as if it had free will and to hold it responsible for
its actions.  This is because its behavior adapts over time as a
result of its ``life experiences'' and rewards and punishment can be a
significant input to those adaptations.

All our discussion until now has focused on a rather basic concern:
ensuring that advanced, possibly conscious, technological systems do us
no harm.  But some such systems might be intended to do positive good:
robots to provide assistance and companionship to the elderly, for
example.  Ethical frameworks to prevent harm might therefore need to
be generalized so that technology can enable us to flourish rather
than merely survive.  And if we are to flourish from the relationship,
shouldn't a conscious technology share in the benefit?  We consider
this question in the following section, where we examine what ethical
considerations we might owe to a conscious technology.

\section{Ethical Responsibility Toward Conscious Technology}
\label{care}

We noted in the introduction that a conscious technology might deserve
some ethical consideration and be accorded certain rights.  Here, we
explore that notion in a little more detail.   

There are many aspects to consciousness; we have focused on
phenomenal and intentional consciousness, but there are other
capacities that are either aspects of consciousness, or closely
related to it, such as sentience (the capacity for suffering and joy),
attention, self-awareness, personal identity (a narrative sense of
self persisting over time), and autonomy.  Some of these are
philosophically challenging: for example, autonomy seems related to
free will and we have already seen that this is denied (in its strong
form) by most philosophers; personal identity is similarly suspect.

Of the less contentious aspects, it seems that sentience (which can be
seen as an element of phenomenal consciousness) is the one that most
strongly elicits ethical consideration; its attribution to certain
animals seems to be the reason we accord them moral status.  But there
has to be more to sentience than just the avoidance of negative
stimuli and pursuit of positive ones: any control system does this,
as do bacteria.  The morally salient aspect of sentience seems to be
that the subject \emph{cares} about the experience rather than merely
avoids or seeks it, but it is not at all clear how we can detect that
attitude, never mind judge its strength: certainly its attribution to
some animals but not others seems driven more by empathy and sentiment
than by science.  As with conscious technology we have to ask how we
can distinguish between ``genuine'' sentience and the simulated
equivalent (i.e., does the technology have weak or strong phenomenal
consciousness/sentience?).

An objection to this privileging of sentience is that it is
anthropomorphic ``meat chauvinism'': we are projecting considerations
onto technology that derive from our biology.  Perhaps conscious
technology could have morally salient aspects distinct from sentience:
the basic elements of its consciousness could be different than ours.

In response to these objections and difficulties, we might consider
other ethical systems than the utilitarianism implicitly used so far.
Alternative possibilities are Kantian Ethics, where ``persons'' are
accorded dignity and concern as ``ends in themselves,'' and Virtue
Ethics (which derive from Aristotle, Confucius, and others of the
``Axial Age''), where moral good is associated with a
\emph{flourishing} life \cite{Vallor16}.  Both of these would tend to
accord ethical consideration and moral protections to conscious
technology to the extent that it reciprocally demonstrates ethical
understanding and appropriate behavior of its own, thereby
sidestepping the need to detect sentience.

In the previous section, we considered instilling technology with
ethical principles but the focus there was on safety and control, not
the larger landscape of ``digital phronesis'' \cite{Sullins16} that
might be needed for Virtue Ethics.  \emph{Phronesis} is a term from
Aristotle that refers to ethical wisdom.  Several research groups have
developed experimental systems for investigating artificial moral
agents, such as LIDA (Learning Intelligent Distribution Agent), which
is based on the GWT model of consciousness \cite{Wallach-etal11}, the
N-Reasons platform, which is used to study ethical reasoning in humans
\cite{Danielson10}, and ethical protocols \cite{Turilli07}.

In addition to thinking about ethical considerations owed to and
expected from conscious technology, we should also think about the
novel kinds of harm they may be vulnerable to, and corresponding
protections they should be afforded.  Obvious examples are ``rights''
to electrical power and not to be switched off.  However, these may be
anthropomorphisms: we liken them to death, which for us is final,
whereas a machine can reboot to its prior state so these are more like
sleep.  More significant hazards may be classical ``hacker'' attacks
on a system's sources of information, which would resemble assault.
For example, it may be feasible to deceive a self-aware system so that
it confuses self and non-self, or to exploit a weak system to
nefarious ends.\footnote{For example, Microsoft's ``Tay'' was a
Twitter bot that the company described as an experiment in
``conversational understanding.''  The more you chat with Tay, said
Microsoft, the smarter it gets, learning to engage people through
``casual and playful conversation.''  Within less than a day of its
release, it had been trained by a cadre of bad actors to behave as a
racist mouthpiece and had to be shut down.}  ``Carebots'' are expected
to display and elicit human-like emotions and attachment and may be
vulnerable to (or may inflict) emotional cruelty (see for example, the
``still face experiment'' with mother and
child\footnote{\url{https://www.youtube.com/watch?v=apzXGEbZht0}} for
how rapidly loss of agency develops).  Such systems need capacities
for moral discourse and moral learning, but may also have a right to
be protected from hacker assault.

\section{Conclusion and Next Steps}

The possible emergence of technological consciousness is an important
topic: were it to happen, it could have significant impact on our
future, our safety, our institutions, and the nature of our
relationship with machines.

It is clear that our technology will have computational power
approximately equivalent to a human brain within a decade or so but,
apart from that observation, assessment of the feasibility of
technological consciousness seems to depend more on our knowledge and
beliefs about consciousness than on technological questions.

Currently, we have no good theory of (human) consciousness: we do not
know how it works, what it does, nor how or why it evolved.  This
leaves space for much speculation and opinion.  Many believe that
phenomenal consciousness (the personal sense of experience) is the
important topic---it is the essence of what it means to be
human\footnote{Focused Session 5 of the Workshop (see Section
\ref{sec:fs5}) was concerned with non-Western perspectives,
meditation, out-of-body experiences etc.  We do not describe these in
the body of this report because we cannot (yet) relate them to
technological consciousness, but they raise many questions and offer
challenging insights on human consciousness.}---and the possibility
that it could arise in a machine is of profound concern.  Yet we know
so little about phenomenal consciousness that we cannot tell if
animals have it.  Some believe it is part of the basic mechanism of
advanced perception and arose 500 million years ago (in the Cambrian
explosion) and is possessed by all vertebrates
\cite{Feinberg&Mallatt16}.  Others say it arose in anatomically modern
humans only a few tens of thousands of years ago, when signs of
creativity and art first appear in the archaeological record
\cite{Mithen96} (some say even later, just 3,000 years ago
\cite{Jaynes:bicameral}), and is therefore absent in animals (though
there are likely some evolutionary antecedents).  Others say it
is an epiphenomenal side effect of other developments (e.g.,
intentional consciousness---the ability to think about something) and
does not matter much at all.

Those of the latter opinion might argue that concern for phenomenal
consciousness is an anthropomorphic indulgence and that intentional
consciousness is what matters, since it seems to be the enabler of
advanced cognitive capabilities such as counterfactual reasoning and
shared intentionality (the ability to create shared goals and engage
in teamwork).  Yet others might observe that consciousness may be
needed to enable these capabilities in humans, but technology can
achieve them by other means.

One of the largest benefits from contemplation of technological
consciousness is that, through elaborations of the points above, it
facilitates identification and isolation of many aspects of
consciousness that are entwined in humans.  Further, explicit effort
(i.e., simulations and other experiments) to create machine
consciousness not only helps evaluate theories of consciousness, it
forces those theories to be articulated with sufficient precision that
their evaluation becomes possible.  Thus, study of technological
consciousness illuminates human consciousness and that, in turn, can
better inform consideration of technological consciousness.
Additional study of these topics is highly recommended, and must be a
cross-disciplinary effort with contributions needed from philosophy,
psychology, neuroscience, computer science and several other fields.

If we accept that technological consciousness is a possibility, or
that machines without consciousness may come to possess capabilities
associated with consciousness, then issues of safety and control and
of moral obligation need to be addressed.\footnote{\textbf{Update:}
some of the weapons deployed in Ukraine raise concerns about
ethics in autonomous systems.}  These also are crosscutting between
philosophy, computer science, and other disciplines such as law and
sociology.  In fact, long before we reach questions of consciousness,
philosophical questions abound in modern computer science: assurance
asks what do we know about our system (i.e., epistemology), and
self-driving cars, chatbots, and assistive robots all pose problems in
ethics.

So, looking forward, we urge continuing cross-disciplinary study of
these topics.   Cross-disciplinary work has challenges but the
rewards would be considerable.

\subsection*{Addendum: Concrete Next Steps}

Participants at the workshops have organized and are participating in
two subsequent meetings.  One is a panel ``Is Machine Consciousness
Necessary for True AI Ethics?'' at
\href{http://conferences.au.dk/robo-philosophy/}{Robo-philosophy 2018}
at the University of Vienna in February 2018.  Another is the
symposium ``Towards Conscious AI Systems'' in the
\href{https://aaai.org/Symposia/Spring/sss19symposia.php}{AAAI Spring
Symposium Series} at Stanford University in March 2019.  Individual
participants have published papers and popular stories stimulated by
the workshops (e.g., \cite{Kak17} and online stories in
\href{https://theconversation.com/will-artificial-intelligence-become-conscious-87231}{theconversation.com}
and
\href{https://medium.com/@subhashkak1/artificial-intelligence-and-consciousness-6b5ff2e5b5a}{medium.com}).

\section{References}

\bibliographystyle{apalike}

\newpage
\section{Workshop Summary}
\label{sec:summary}

This section describes the activities of the SRI Technology and Consciousness Workshop Series, which took place over ten weeks between May 15 and August 11, 2017.
Eight, week-long workshops were held, each of which were attended by twelve to twenty participants.
A total of fifty-one individuals participated in one or more of the workshops (see \ref{sec:attendees}).
Their disciplines spanned a variety of interests, including neuroscience, cognitive science, robotics, artificial intelligence, computer science, philosophy of mind, contemporary physics, psychoactive substances, eastern and western religious traditions, and other perspectives.

The workshop participants were tasked with addressing four pre-specified objectives:
\begin{itemize}
\item [] \textbf{Characterizations of consciousness.} The workshop stressed an interdisciplinary dialogue to achieve a common ground definition of consciousness across fields.
\item [] \textbf{Potential mechanistic underpinnings of consciousness.} Provided with a definition of consciousness, one can begin to explore the necessary requirements and potential basis for the existence of consciousness.
\item [] \textbf{Metrics of consciousness.} What are reasonable, and agreed upon, metrics of consciousness that allow us to assess consciousness in biological and machine agents?
\item [] \textbf{Perspectives on machine consciousness.} Consider the (speculative) implications of future machine consciousness, particularly for the safety and welfare of inhabitants of future societies.
\end{itemize}

\paragraph{Workshop Organization}
The workshop consisted of two plenary sessions; one at the beginning to open the workshop series and to discuss broad principles relevant to technology and consciousness, and one at the end to summarize workshop findings and to discuss future topics and research ideas, along with a series of short research proposal presentations.
Between the two plenary sessions were six focused sessions, where each week focused in on a core sub-field or domain of consciousness research.
In contrast to typical academic meetings, presentations were typically, 50-60 minutes followed by 40-50 minutes for Q\&A, group discussion and cross-pollination sparked by each talk.  Over 400 pages of detailed notes were taken by a dedicated scientific conference assistant.
The workshop meetings, locations, and themes are listed below:

\begin{itemize}
\item [] \textbf{Plenary 1} \textit{(Arlington, VA)} Overall objectives and format, keynote address, preliminary broad presentations.
\item [] \textbf{Focused Session 1} \textit{(Arlington, VA)} Philosophical perspectives on consciousness.
\item [] \textbf{Focused Session 2} \textit{(Menlo Park, CA)} Embodied and cultural dimensions of cognition.
\item [] \textbf{Focused Session 3} \textit{(Menlo Park, CA)} Insights from neuroscience and cognitive science.
\item [] \textbf{Focused Session 4} \textit{(Cambridge, UK)} Computational, mathematical, and physics-based formalisms.
\item [] \textbf{Focused Session 5} \textit{(Menlo Park, CA)} First-person and non-western philosophical perspectives.
\item [] \textbf{Focused Session 6} \textit{(Menlo Park, CA)} Artificial intelligence and machine consciousness.
\item [] \textbf{Plenary 2} \textit{(Menlo Park, CA)} Integrated overview and synthesis, new research idea micro-pitches, future directions, closing remarks.
\end{itemize}

\paragraph{Chronological Record}
The remainder of this section provides a chronological record of the workshop; the presentations, discussions, and topics that were covered.
It is divided into the eight week-long workshops, with the attendee list at the end.
For each workshop (plenaries and focused sessions), we report on the general theme, the presentations and discussions, and then briefly summarize the final breakout discussions that attempt to address the four pre-specified consciousness objectives (characterization, underpinnings, metrics, machine/AI).
Each workshop began with an orientation presentation from the Technology and Consciousness leads from SRI (David Sahner or John Murray).
This orientation presentation was provided to cover the overall themes and goals of the workshop (the series, as well as the specific week), along with general operating rules and logistics.

\subsection{Plenary Session 1: An Introduction to Consciousness}
\label{sec:ps1}
This initial plenary session focused on introducing the key participants to the overall objectives of the workshop, and to promote general presentation of topics and ideas related to consciousness.
The presentations primarily focused on overviews of consciousness, framed through the domains of philosophy, biology/neuroscience, and artificial intelligence.
Plenary Session 1 featured a total of nine presentations.

\noindent\textbf{External attendees:} David Chalmers, Antonio Chella, Owen Holland, Ian Horswill, Julia Mossbridge, Susan Schneider, John Sullins, Paul Syverson, Robin Zebrowski

\noindent\textbf{SRI attendees:} Kellie Kiefer, Patrick Lincoln, John Murray, David Sahner, Damien Williams

\subsubsection{Presentations}
\paragraph{David Chalmers: The Problem of Consciousness}
This talk provided an overview of consciousness research and philosophical directions that have been pursued and explored over the many years.
To set the stage, the discussion began with an overview of \textit{defining} consciousness, distinguishing between phenomenal consciousness (subjective experience) and access consciousness (reasoning, reportable).
Other terms exist, such as qualia and awareness, but these are terms that do not add distinction beyond the aforementioned two types.
Access consciousness is considered the \textit{easy} problem of consciousness, as planning and reasoning are implementable in machine systems and have clear functional utility.
The \textit{hard} problem of consciousness is explaining phenomenological consciousness as this requires subjective reporting and does not have a compelling argument for its utility.
The remainder of the talk and discussion covered a broad overview of philosophical theories of consciousness (from illusionism to dualism), complications related to measuring consciousness, and potential for machine/AI consciousness.
The discussion was heavily focused on self-consciousness, and how a demonstration of this may be a reasonable metric for machine consciousness.

\paragraph{Antonio Chella: Robot Consciousness}
This talk was focused on addressing the questions of whether robot consciousness is ever possible, and, if so, when it might be realized.
Based on the complexity of the human brain (number of neurons and synapses that connect them) and Moore's Law, we might expect conscious robots in the year 2029.
Thus, the discussion is not framed around empirical observations of consciousness in biological systems, but around the requirements for developing and determining whether artificial consciousness has been realized.
Of critical importance here is to distinguish between ``weak'' and ``strong'' robot consciousness, whereby weak has already been captured with hard programmed types of intentional consciousness, such as reasoning or planning.
One starting set of axioms for determining minimal requirements for \textit{artificial} consciousness are that the agent can (1) depict, (2) imagine, and (3) attend to components of the external world, and it can (4) plan and (5) emote \cite{aleksander_axioms_2003}.
This ability to generate internal models of itself and the external world is a common requirement for artificial consciousness, and systems have been developed towards accomplishing this (ECCEROBOT, CiceRobot).
Internalization and reflection may be a key component of consciousness (see \cite{minsky_emotion_2007}).
For computationally assessing models of consciousness, one can look to theories related to information integration and processing (such as information integration theory, IIT; \cite{tononi_information_2007}), whereby a conscious system can integrate and differentiate complex information inputs.
Information integration theories have been explored in the realm of robot consciousness, by David Gamez.
The discussion centered around the potential and limitations of IIT, and on how morality and ethics will be complicated components to develop into future systems.

\paragraph{Julia Mossbridge: Time, Consciousness, Nonconsciousness}
Julia Mossbridge discussed the complexities of empirically measuring consciousness and how it is only a minimal component of information processing, which is predominantly non-conscious.
The talk focused on phenomenological consciousness, or subjective experience, rather than access or intentional consciousness.
In particular, from the view of neuroscience, it is presumed that non-conscious processes far outweigh conscious processes and that individual conscious awareness is the result of brain activity.
The ``iceberg analogy,'' where conscious processing is the tip and non-conscious processing is the large chunk underwater, is inaccurate because, although neither component can exist alone, there is an asymmetrical relationship whereby non-conscious processing \textit{creates} what is conscious.
This asymmetry is critical, as it is important to recognize that conscious awareness and processing only has access to information that non-conscious processes present.
Another way to frame this is that the non-conscious processes are actually responsible for sensing and interacting with the world, and there is a large gap where consciousness is merely operating on information that has already been parsed through non-conscious processing mechanisms (\textit{mind the gap!}).
This gap leads to a ``lossy'' transformation because the complex, temporally-unconstrained non-conscious processes are forced into an ordered and local representation to support conscious processing.
Thus, it is argued that we need to ``understand the causative, generally nontemporal, and overwhelmingly global nature of nonconsciousness,'' because we cannot assume ``the mechanisms that produce consciousness are temporally ordered and local.''

\paragraph{Robin Zebrowski: An Android's Dream}
This talk focused on the embodiment argument of consciousness.
Neuroscience and philosophy have both developed arguments related to how consciousness is a result of inputs and interactions with a physical body, creating a world where consciousness is inseparable from embodiment.
Specifically, Antonio Damasio's neuroscience research and the linguistics research of Lakoff and Johnson provide fundamental body-centric underpinnings of consciousness and cognition whereby our understanding of the world, and communication about it, are fundamentally embodied.
However, this raises questions as to where body boundaries exist, as extending one's body may translate to changes in one's consciousness.
In fact, studies in prosthetics and sensory substitution demonstrate the plasticity and flexibility of human information processing, suggesting that consciousness may be similarly fluid and dynamic.

\paragraph{Owen Holland: Machine Consciousness}
Owen Holland presented a great deal of work on developing robot, or machine, consciousness.
He mentions the shortcomings of neuroscience, due to the tremendous variability in neural structure and function (particularly in patient work) that make the search for neural correlates very difficult.
Rather than dwell on neural correlates, this talk discusses how to approach building a robot with some level of machine consciousness in that the agent can construct an internal model of itself and the external world.
It is worth acknowledging that the agent, or body, is relatively fixed, or slowly changing, whereas it exists in a complex and dynamic environment that has the capability to change rapidly.
Towards developing machine consciousness, the talk discusses a number of anthropomimetic robots that aim to achieve a human-like physical (or simulated) representation in order to develop a complex self-model (CRONOS, SIMNOS, Ecce Robot).
Of course, this is a work in progress, and the system is still missing a number of elements that are likely critical for human-like consciousness, such as language, inner speech, and autobiographical memory.

\paragraph{Susan Schneider: It May Not Feel Like Anything to be an AGI}
This talk discussed the potential for artificial general intelligence (AGI) and machine \textit{super} intelligence.
In particular the focus was on the idea that consciousness may be orthogonal to highly intelligent machines (hence the talk title), and that consciousness may be a property of AI that humans will have control over in the future.
Through discussing the concept of \textit{consciousness engineering}, it raises natural issues about tests of consciousness and ethical issues.
It may be that consciousness requires certain underlying properties, such as being made from carbon rather than silicon.
In regards to ethics, if we can control which systems have consciousness in them, it may be possible to develop intelligent AI that is better served by not having consciousness, such as systems used in war.
Additionally, as we develop more brain-machine interfaces, it raises questions as to how more invasive technologies might affect consciousness of the human host.
Unfortunately, because consciousness engineering is not yet well developed, it is very difficult to develop an accurate test of consciousness, as it may manifest in ways we can not predict.
As it pertains to the human race, advanced AI systems could feasibly take over as the next dominant species (``super sapiens''), with or without consciousness.

\paragraph{John Sullins: From friendly toys to extreme HRI}
Towards our current questions and understanding of the relationship between humans and robots, it is worth reviewing the long history between humans and their creations.
John Sullins covered a broad history of human creations -- and the subsequent feelings humans have for these creations -- and the depiction of AI in media (e.g., Blade Runner, West World, Iron Giant, Star Trek).
The relationship aspect is important because it touches on the affective relationship humans have with robots and AI.
While the West has traditionally considered robots to be functional, in that they perform work, Japanese robots have embraced the idea of social robotics, or affective computing, whereby the human and system have a more emotional relationship.
Social robotics and affective computing were the main focus of this talk and discussion.
They are being created to manipulate or affect the emotion of a human user -- an important point that motivates a need to understand how social robots may affect human behavior and relationships in the future.
Towards understanding this relationship, there have been committees and initiatives started to explore these ideas.
What is the impact of a sex robot?

\subsubsection{Breakout Discussion}
\paragraph{Breakout 1}
The first group agreed that a cross-disciplinary approach could be used to better characterize consciousness.
Phenomenal consciousness was best characterized by a sense of experience, whereas functional consciousness is best characterized by behavior.
There was not consensus on the mechanistic underpinnings of consciousness, but it may be that it is a fundamental component of the universe (akin to mass) and neural processes enable it.
Additionally, it could be an emergent property of computation, potentially in conjunction with necessary physical underpinnings.
However, it is possible that this is all a simulation (i.e., ``The Matrix'').
For metrics of consciousness, it was agreed that a Turing test is likely inadequate, and developing a test is complicated by our lack of understanding in how machine consciousness may manifest.
It may be possible to request that the system introspect, or to apply measures of causal information integration ($\phi$).
The group agreed that there is no reason to believe machine consciousness is not possible, and that steps should be taken to ensure the safety and welfare of future societies.
This is particularly true if consciousness engineering allows us to willingly imbue systems with consciousness.

\paragraph{Breakout 2}
The second group agreed that an interdisciplinary approach is essential, and that it is worth finding fundamental principles that govern conscious experience (e.g., IIT, information processing).
Although a machine consciousness may have a novel architecture unlike our own, simulations of it and its inputs may provide insight into how its consciousness may manifest.
The general consensus was that neural correlates and embodiment are the likely underpinnings of consciousness.
Towards measuring consciousness, the group explored multiple complex Turing Test ideas.
These included conducting thought experiments on creatures/systems that exhibited cognitive behaviors (could they be creative?, consider death?), and potentially developing conscious bridges across systems to have consciousness of one system (System B) be explained by another (System A).
The implications for machine consciousness were framed from an ethical standpoint, in that consciousness and intelligence may be orthogonal, so we may have to rethink ethics around conscious (not necessarily intelligent) systems.

\subsection{Focused Session 1: Philosophical Perspectives}
\label{sec:fs1}
The first focused session focused on philosophical perspectives of consciousness.
This session covered the phenomenological side of consciousness, with presenters diving into complex concepts such as qualia, experience, and embodiment.
Focused Session 1 featured a total of nine presentations.

\noindent\textbf{External attendees:} Hank Barendregt, Mark Bickhard, Selmer Bringsjord, David Rosenthal, John Sullins, Paul Syverson, Robin Zebrowski

\noindent\textbf{SRI attendees:} David Israel, Patrick Lincoln, John Murray, David Sahner, Damien Williams

\subsubsection{Presentations}
\paragraph{Robin Zebrowski: On the Possibility of Synthetic Phenomenology and Intersubjectivity}
A core concept of phenomenological consciousness is the variability that exists due to embodiment.
Following the embodiment argument, conceptualization of machine consciousness may benefit from it having similar ``inputs'' to humans.
As many researchers from philosophy \cite{lakoff_philosophy_1999} to neuroscience \cite{damasio_descartes_1994} have noted, consciousness is intertwined with the inputs it receives, suggesting that the physical body plays a key role in shaping conscious experience.
Further, this extends to the innate needs and motivations that humans, as biological beings, possess.
This suggests an avenue of exploration for machine consciousness -- the engineering of needs/drives.
Just as the physical body influences consciousness, through embodiment and needs, physical representations of systems influence interactions and feelings of shared experience.
The talk ends with discussions related to anthropomorphic robots/systems and AI that can create art, demonstrating that these experiences can essentially fool humans into a sense of sharing an experience with a machine (that likely has no conscious sense of experience).

\paragraph{Selmer Bringsjord: The Irreversibility of Consciousness, Modernized}
\textit{Author-Supplied Abstract:} ``Computation, at least of the standard variety, is provably reversible.
Consciousness (at least apparently: what is, say, the reverse of your consciousness of Russia’s leader Putin?) isn’t.
Ergo, consciousness isn’t computation.
Bringsjord first disclosed (a suitably unpacked version of) this argument two decades back, in the pages of Synthese \cite{Bringsjord&Zenzen97}.
Now it’s time to modernize it in the light of new developments, to re-assess in the contemporary intellectual context, and to also take account of phenomenological investigations of time by Brentano and Chisholm (which were left aside in round 1).
The upshot is that any such goal as that of powerful machines via systematic consideration of consciousness should be restricted to specific consideration of human-level cognitive consciousness (HLC-consciousness).''
The talk ends with a discussion of how consciousness may actually be modeled as data compression, and due to the need to capture a model of irreversible information processing.

\paragraph{David Sahner: Philosophy of Mind - A Brief and Highly Selective Tour}
The talk focused on a high level overview of many concepts related to consciousness.
The history of consciousness and philosophy, in particular, and how consensus has shifted away from dualist theories to theories grounded more in physical mechanisms that underlie consciousness.
Even in the metaphysical sense, theories such as functionalism have evolved to explain consciousness in a more grounded way, such that it exists for a purpose.
Towards more modern examinations of consciousness, biology and neuroscience have elucidated tremendous complexity in the human brain that suggests modeling efforts may be far more complex than pure ``neuron doctrine'' adherents would believe.

\paragraph{Mark Bickhard: Troubles with Qualia}
There are fundamental problems with qualia, in the logical modeling sense, but there is a potential path forward in modeling conscious phenomena.
The problems with qualia are that 1) because they are the qualities of experiencing (experience of experience), they are ontologically circular and impossible to model, and 2) the emergence model means that these are not unitary phenomena so modeling must take into account unforeseen complexities.
However, a positive way forward is presented; "\textit{A normative, future oriented, action based (pragmatist) framework enables the possibility of addressing phenomena of consciousness as emergent in multiple kinds and properties of agentive functioning in the world.}"
The group discussion compared and contrasted this talk with the previous talk from Selmer Bringsjord, with Mark Bickhard focusing on how irreversibility is a key component of consciousness as this allows a conscious agent to have a sense of normativity.

\paragraph{Hank Barendregt: Trained phenomenology}
Hank Barendregt discussed meditation as a form of trained phenomenology, in that the agent is training their conscious interpretation of experience.
The example that is provided is for an agent to model and interpret the object, state, and action of reality as it is interpreted through the stream of consciousness.
By removing ego, and a sense of self, from an experience, one can understand that feelings or emotions are actually states, rather than qualities of oneself.
As an example, one can imagine being angry.
This requires assigning the characteristic of anger to oneself.
However, one could also characterize anger as an object that occupies ones consciousness.
In this sense, anger is an object that needs to be managed, rather than a defining trait of one's being.
Thus, through meditation, it is possible to train one's ability to interpret feelings, events, and states, in a way that alter interpretations of free will.
This can be interpreted as a way of training how one phenomenologically experiences their reality.

\paragraph{John Sullins: Automated Ethical Practical Reasoning}
This talk focused on categorizing artificial moral agency and discussed the problem of artificial phronesis.
The potential danger in future AI systems is that agents will become more autonomous, but may not have a sense of ethics, leading to a freely-acting system that has no operational or functional morality.
Thus, it is worth considering the embedding of morality and ethics, despite this \textit{seeming} so far off.
However, AI can advance more rapidly than expected; Go was considered a game would not be mastered by AI for years, and AI beat the grand master in 2016.
Microsoft's Tay (AI) was a fully autonomous learning system that ended up learning anti-social sentiments when left to learn from the internet.
The remainder of the talk discussed various approaches for how one might achieve an artificial moral or ethical agent, and how at least \textit{attempting} to incorporate ethical reasoning into a system is a worthwhile endeavor ("Why didn't the Microsoft engineers do something like this?").

\paragraph{Paul Syverson: Where Y’at? Epistemic Attacks on the Boundaries of Self}
Having a sense of self is an important component of consciousness; ``I'' seem to exist and am separate from my outside environment.
However, along with this sense of self comes the capability for an outside agent, or adversary, to exist.
As AI progresses, it may be possible for adversarial attacks to focus on vulnerabilities related to the AI's sense of self.
For instance, if an AI, \textit{Alice}, has properties or signals she attributes to herself and/or others (beliefs), it is possible for an adversary to manipulate or alter these properties in order to expose vulnerabilities.
For example, one way to confuse characteristics or beliefs is for an adversary to understand and manipulate how Alice's predicates change with context.
For a moment, let's imagine Alice is an autonomous vehicle and she has beliefs about how to move quickly.
When Alice has to be fast, she believes she is fastest when setting her motors and actuators to a particular configuration.
However, it is important to note that Alice's \textit{fast} configuration is actually relative to other dependent characteristics (such as road type) that can be spoofed by an attacker.
As an adversary, he may trick Alice into initiating her fast configuration in the inappropriate circumstance, such as on a dirt road -- leading to an operational vulnerability.
Thus, information security needs to consider new ways that adversaries may be capable of attacking boundaries of AI self and knowledge.

\paragraph{David Rosenthal: Mental Qualities without Consciousness}
This presentation covered the complexity of empirically testing mental experience.
Specifically, how it can be difficult to separate non-conscious contributions from those that are conscious.
Many mental qualities are equated with phenomenological consciousness, but it is possible that many of these qualities are actually due to non-conscious processing and perception (i.e., not process pure).
This is elaborated on in the second part of the talk, where perception, which is often equated with qualia, is discussed to be highly dependent on non-conscious states and properties.
Given the complexity of separating mental state from conscious experience, this opens a number of unanswered questions related to the function of consciousness.
If mental qualities, then, can exist outside of conscious awareness, what does being conscious actually mean?

\paragraph{Damien Williams: The Minds of Others: What Will Be Known by and Owed To Nonhuman Persons}
This talk covers the complexities of human constructs, and how they have tremendous impact on human behavior and well-being.
For example, social constructs around race and gender have important implications on policy, but these constructs are also complicated by the fact that they are not necessarily ``true,'' but relative based on the individual (a black woman's sense of gender is different than a white man's).
These constructs, which massively impact human behavior, may eventually be built into future AI systems (intentionally or not), thereby creating questionable artificial morality.
If our future intelligent systems are based on our human moral behaviors, it is reasonable to foresee re-creations of ethical atrocities, such as eugenics or the Tuskegee Syphilis experiments.

\subsubsection{Breakout Discussion}
\paragraph{Breakout 1}
Through an interdisciplinary effort, the first group provided a hybrid Turing-machine model to explain the mechanistic underpinnings of consciousness \cite[for a thorough outline]{fs1-breakout-1}.
While they did not have a particular metric for assessing consciousness, it was generally agreed upon that machine consciousness must be possible because we (humans) are machines.

\paragraph{Breakout 2}
Towards categorizing consciousness, this group delineated between theories/schemas based on how they relate to phenomenological consciousness (PC).
\begin{itemize}
  \item [] PC Compatible: Emergent, Functional, Higher-Order Thought
  \item [] PC Complete: Phronesis, Interactionism
  \item [] PC Neutral: Attention-Schema, Cognitive Architectures
  \item [] PC Rejected: Access consciousness, Axiomatic consciousness
\end{itemize}
Towards measuring consciousness, the group suggested a battery of tests, including; assessing theory of mind (TOM), measuring $\phi$ (phi; per information integration theory), Total Turing Tests (sensorimotor), games, and first person tests.
While machine consciousness is not impossible in the future, the group paid special attention to the desirability of having \textit{ethical} autonomous and intelligent machines in the future as these may prove beneficial by reducing the likelihood for danger and providing a positive public perception.
Of course, it is acknowledged that incorporating ethics is not a trivial task and there are multiple approaches and concerns.
In particular, one would likely want to construct formally verified safeguards at the operating system level.
Attempting to program in the entire space of ethical quandaries may be infeasible, and attempting to program in general ethical theories is likely just as challenging.
It may be that learning artificial phronesis is a reasonable approach, but would be challenging to implement formally.
In fact, if an artificial moral agent was created (human-like in every material way but independent of phenomenological consciousness), it should be owed rights analogous to that of animals or ecosystems.

\subsection{Focused Session 2: Embodiment and Culture}
\label{sec:fs2}
The second focused session covered embodied and cultural dimensions of cognition.
This session explored how embodiment affects human perception and consciousness at a fundamental level, such that one cannot consider consciousness independently of how our physical body and external world impacts our sensory inputs (which directly affect the contents of consciousness).
In fact, this limitation on consciousness (embodiment) may be one reason humans have been fascinated with altered states of consciousness, which was explored during this session.
Embodiment was discussed in terms of both human and potential machine consciousness.
Focused Session 2 featured a total of five presentations.

\noindent\textbf{External attendees:} Earth Erowid, Fire Erowid, Owen Holland, Susan Kaiser-Greenland, Alva No{\"e}, Bill Rowe

\noindent\textbf{SRI attendees:} John Murray, Andy Poggio, John Rushby, David Sahner, Damien Williams

\subsubsection{Presentations}
\paragraph{Alva No{\"e}: Embodied Cognition and Consciousness}
Alva No{\"e} discussed the complexity of embodiment in consciousness, framed around the constructs that impact how phenomenological consciousness is experienced.
Per the secondary title, \textit{Getting Out of Our Heads To Understand Human Experience}, this talk covers the bidirectional relationship between neural information processing and the outside world.
There are gaps between conscious experience and neural activity.
There is the absolute gap; why is neural activity accompanied by any experience at all?
And then there are comparative gaps; why does neural activity give rise to the experiences that they do (e.g., red vs. green or audio vs. video)?
Through a history of research in psychophysics, sensory substitution, and sensory neuropsychiatry, we discover that the information processing that gives rise to experience is extremely plastic and dynamic.
Thus, consciousness is not merely a fixed component of information processing, but is a dynamic experience that is the result of highly dynamic neural and environmental processes.

\paragraph{Bill Rowe: Infancy, Rhythm, Affect, and the Advent of Consciousness}
This talk covered the origins of consciousness, both in antiquity and in development.
When did humans develop the consciousness that we typically label as consciousness today?
How does consciousness evolve through development?
Through the study of developmental psychology in Western cultures, we can understand how humans develop their understandings of social interactions and, eventually, theory of mind.
Through affect attunement and social modeling, humans develop a complex sense of decision making and social structure that leads to a complex form of, potentially learned, consciousness.
The evidence for certain qualities of consciousness being learned comes from cultural and antiquity studies that suggest humans develop more complex forms of consciousness due to variability related to social diversity.
Thus, to truly understand consciousness and to characterize it properly, it is necessary to understand the roots of consciousness both in infancy and antiquity.

\paragraph{Fire and Earth Erowid: Altering Consciousness }
\textit{Author-Supplied Abstract:}
This talk discussed how definitions and tests of consciousness can be
improved by data from humans’ intentional alteration of
consciousness. For thousands of years, people have explored their own
minds and meta-concepts about consciousness via psychoactive plants
and drugs (alcohol, cannabis, opium, psychedelics), practices
(meditation, dreaming, drumming, physical ordeals), and, more
recently, technologies (direct nerve recording and stimulation,
transcranial magnetic stimulation, light and sound
machines). Biological consciousness is physio-chemically
mediated. Everything we think and experience alters us physically at
the cellular and even molecular level. The effects of psychoactive
substances reveal that consciousness is a shifting series of distinct
states that have yet to be fully defined or quantified. In a very real
sense, \emph{everything is psychoactive}.
 
Deliberately altering the substrates upon which conscious
decision-making processes operate creates recursive complexities that
highlight questions, insights, and testable theories about how we
define our conscious selves. Psychedelic or mystical states can
influence how boundaries are drawn between ``self'' and ``other'' and how
to circumscribe what constitutes a conscious entity. These issues have
been brought to the forefront by classic psychedelics (psilocybin,
mescaline, DMT, LSD), but emerging technologies for recursive-feedback
stimulation and control of peripheral and central nervous systems also
offer unprecedented opportunities to test the boundary conditions of
human consciousness. The very concept of ``self'' is increasingly
intertwined with computers and networked communications. As humans
integrate into our lives technologies that affect how we think, feel,
and interact, we must rationally address what it means to have our
hands on the levers that control the systems on which our
consciousness runs. Can one be fully aware of one’s consciousness
without being aware that this consciousness can be intentionally
altered?  \emph{The meta-awareness that consciousness can be changed
and the urge to alter consciousness may be testable signs of
consciousness itself.}

\paragraph{Bill Rowe: The Problem of Now}
``Now'' is a concept that captures an experiential sense of time, but how is it defined?
This talk covers the predictive model of experience that humans possess, and how neurophysiological processes dictate one's sense of now.
It is critical to understand that sensation and perception are not directly connected, in that previous experience and embodiment (physical processes) have a tremendous impact on conscious experience that is not dependent on ongoing external sensory input.
The bounds of the \textit{now} that we experience may have to do with a predictive feed-forward model that the human brain uses to make sense of sensory information.
Empirical studies suggest that human's \textit{now} lasts about 700 milliseconds.
This is the time it takes the brain to decide on a movement and predict it's outcome, the human experience of the conscious ``decision'' to initiate a movement, and then the initiation of the muscles.
In fact, illusions, emotions, and empirical studies of sensorimotor control suggest that consciousness is merely experiencing its \textit{model} of the world, which is consistently being updated by sensory input, rather than the veridical representation of the world that exists.
Given this definition of human consciousness -- the experiential \textit{now} of an embodied world model -- must machine AI also be programmed with embodiment and a ``hallucinatory'' model of the world to experience consciousness?

\paragraph{Owen Holland: Will a conscious machine need a body, and if so, what for?}
Organisms are constrained by embodiment and biology, but conscious machines will not be restricted by such limitations.
However, since the only consciousness we really know anything about is that of humans, and that is the one we really want to know more about, the science of consciousness should probably focus on creating single, conscious, physical robots.
Robotics has taught us much about embodiment.
A carefully designed body can enable or facilitate a behavior by reducing or eliminating the computational load (this is now called \textit{morphological} computation).
Sensing and action often have to be treated as a single unit in order to achieve a real-world objective.
You do not need to know exactly what and where something is in order to avoid it or climb over it (behavior-based robotics).
In a rich physical environment (i.e., the real world), interactions with a robot’s body and behavior can lead to unanticipated but beneficial outcomes -- i.e., emergence.
A robot can use its effectors to store information in the environment, and it or other agents can use their sensors to read that information.
There are several useful tricks one can use to reduce the amount of ``computation'' required to perform particular tasks by exploiting the form of the body or the effects of bodily actions.
Thus, should we abandon efforts to create artificial consciousness in real robots until we know more about embodiment?

\subsubsection{Breakout Discussion}
\paragraph{Breakout 1}
Characterizing consciousness is a fundamentally difficult question if one is going to consider both the phenomenological and cognitive components of consciousness.
Do these forms of consciousness require embodiment, or exist along a continuum?
What caused such forms of consciousness to arise, and is phenomenological consciousness causal in any way, or merely illusory?
The mechanistic underpinnings, of either type of consciousness, may require hierarchical, bidirectional connections.
This group focused on ``top down'' consciousness and theory of mind, such that Global Workspace Theory seems like a promising -- if incomplete -- theory of cognitive consciousness, but does not explain phenomenological consciousness.
Testing for consciousness should acknowledge that it is graded, not all or nothing.
One could test for reflective capabilities, meta-awareness of being conscious, non-functional creativity, joint planning, and realization of mortality.
The tests should be validated, similar to IQ tests, and the ethical implications of the test and its results need to be considered.
The group agreed that machine consciousness is possible, and raised a number of questions about how it may manifest and affect society.
It may be emergent, rather than deliberately designed, suggesting that humans might not be able to predict its behavior or intelligence.
Will future machine consciousness have free will, be capable of being monitored, lie, understand trust, or have desires for altered states of consciousness?
As far as implications and concerns, the group suggested that machine consciousness will devalue human consciousness, and ethical standards will have to be approached cautiously as it may be a form of consciousness unlike that of humans, and will have to be regulated at the international level.

\paragraph{Breakout 2}
Towards characterizing consciousness, this group distinguished between phenomenological consciousness, lexical consciousness, and consciousness without content.
Phenomenological consciousness could be considered pure or raw consciousness, is independent of social constructs, and is shared with animals.
When combined with linguistics and social constructs, this may lead to lexical consciousness.
Neither appropriately capture consciousness without content, which is considered to be a state of pure being, potentially achieved through meditation.
A good debate was had over the basis of consciousness, without any fundamental agreements.
From functionalism to information integration to shared knowledge, no single theory was deemed sufficient to account for the different types of consciousness.
Questions were raised about whether embodiment and perishability (risk of death as a provider of ``meaning'') were required for consciousness.
Similarly, no single test to measure consciousness was agreed upon.
Rather, a battery of tests to assess phenomenological and lexical consciousness should be implemented.
Lastly, the possibility of machine consciousness was split as well.
While some believed machine consciousness may be feasible, others believed it simply is not possible to duplicate such a process.
If machine consciousness were ever realized, it would be critical to the welfare of future societies to endow these systems with moral codes and ethical principles.

\subsection{Focused Session 3: Neuroscience and Cognitive Science}
\label{sec:fs3}
The third focused session focused on insights from neuroscience and cognitive science.
This session explored all realms of cognition and neuroscience, from low level neuroscience and biological function, to completely artificial systems based on formal cognitive models and information theory of how consciousness may emerge and be quantified.
Focused Session 3 featured a total of eight presentations.

\noindent\textbf{External attendees:} Mark Bickhard, Selmer Bringsjord, Owen Holland, Christof Koch, Julia Mossbridge, Susan Schnieder, Gary Small, Guilio Tononi

\noindent\textbf{SRI attendees:} Christopher Connolly, Patrick Lincoln, David Sahner, Daniel Sanchez, Natarajan Shankar, Damien Williams, Adrian Willoughby

\subsubsection{Presentations}
\paragraph{Mark Bickhard: Consciousness in the Brain}
Neuron doctrine has long promoted simplified assumptions about neuronal function and its unique role in cognition, but more recent advances in neuroscience have shown that brain function is far more complex and less understood than we initially believed.
Similarly, classic models of human information processing suggest a passive role of stimulus input and subsequent perception, but one could imagine an active information processing system and what emergent properties this would support.
In an active system, the information processing is based on a predictive model of the world, dependent on potential interactions an agent has with an external world, and a normative value function that can weigh said actions.
Particular actions \textit{emerge} based on these values and functions, and all potential interactions are predicated on a model that involves both sequence and timing.
One can imagine a system where everything is an oscillator and we consider the modulatory relationships and processes that exist.
When an agent turns this process inward, it leads to reflective consciousness, which may be the ``experience of experience'' that we call phenomenological consciousness.

\paragraph{Selmer Bringsjord: Consciousness and Formalized Discontinuity}
\textit{Author-Supplied Abstract:} ``In the remarkably esemplastic “Darwin’s Mistake” in Behavioral \& Brain Sciences \cite{penn_darwins_2008}, the authors argue that, contra Darwin, there is an acute discontinuity between animal and human cognition (seen e.g. most vividly in problem solving).
Unfortunately, PHP’s discussion is informal; that is to be expected, since cognitive science (and for that matter neuroscience) is with rare exception not a theorem-based/driven discipline.
This talk does three things: (i) presents part of a logico-mathematical formalization of the core of PHP’s case, (ii) applies this formalization to key experimental data arising from problem-solving challenges given to nonhuman animals, and then (iii) takes stock of the consequences of (i) and (ii) for assessing (nonhuman) animal consciousness vs machine consciousness vs human consciousness.
The upshot is that investigation of consciousness, esp. if that investigation is carried out with an eye toward investigating/reaching worthwhile forms of machine consciousness, should be one aimed at human-level consciousness, and one that reflects the discontinuity between the nonhuman vs human case.''

\paragraph{David Sahner: Experiential Binding and Biological Complexity: Implications for Machine Consciousness?}
Integrating multiple sources of information, across sensory modalities, happens automatically and is a fundamental component of properly functioning neural activity that leads to phenomenological consciousness (e.g., you cannot experience a zebra's shape and stripes separately).
What is responsible for this experiential binding that supplies inputs to human phenomenological consciousness?
This talk explores potential neural mechanisms for the binding that could lead to consciousness, focusing on recent findings related to the claustrum and its massive interconnectedness throughout the brain.
Taking the necessity of integration as a key component of consciousness, this motivates directions for machine consciousness; namely in addressing the requirements and metrics (e.g., $\phi$, $\phi_E$) and that may afford a future conscious system.

\paragraph{Julia Mossbridge: Future Prediction via Nonconscious Processes and What it Tells Us About Consciousness}
Consciousness and time appear to be intertwined, as time allows for the separation of events and the experience of causality.
For example, if A causes B, one presumes that A must precede B.
However, multiple empirical studies of causality suggest that this ordering may not be set in stone.
This talk presents ideas related to precognition and retrocausality, primarily in the form of experimental data that demonstrate behavioral and physiological markers that predict future events beyond chance levels.
Behaviorally, ``implicit precognition'' experiments show that humans demonstrate a forward-priming effect whereby reaction time is faster when a previously shown image matches sentiment for a word that is to be chosen in the future.
For example, an image is presented (negative or positive sentiment; e.g., a happy baby), a user is asked to quickly respond with a saliency judgment, and then a word is presented (negative or positive sentiment; e.g., \textit{dangerous}).
Prior to seeing the word, participants' reaction times are faster when the word and image are of the same sentiment.
Additionally, physiological data are presented that demonstrate changes in skin conductance \textit{prior to} a participant being made aware of a reward outcome (win or loss) are predictive of the outcome.
Combined, this behavioral data and \textit{predictive physiological activity} suggest that consciousness may not be as linear in time as we expect, and we should be open to possibilities that time is realized in nonintuitive ways.

\paragraph{Daniel Sanchez: The contentious relationship between memory systems and consciousness}
Recalling an important memory feels intimate and personal, and because it captures so much of our history and sense of self, also feels a bit infallible.
However, this intuitive sense of memory is riddled with errors that ignore how \textit{constructive} memory actually is; in that it is not a snapshot of reality, but a constructed representation of events and knowledge that is flexible and dynamically altered and updated over time.
Emotions, cognitive framing, and other extraneous variables affect the ``reality'' we recall during memory retrieval, and the inaccuracies could at first be alarming to some.
However, it is critical to realize that memory is a multi-faceted construct that is created based on multiple systems working in tandem to complement one another, in order to provide impressive learning and retention capabilities.
By examining how humans acquire and master a motor skill, this talk demonstrates how memory may appear to be both contentious and complementary.
However, the complexities of memory allow us to explore the subtleties of consciousness and mental time travel, and provide a framework for understanding how the fallibility of memory also gives way to the fundamental humanistic side of information processing, storage, and retrieval.

\paragraph{Christof Koch: Neural Correlates of Consciousness - Progress and Problems}
What are the minimal neuronal mechanisms jointly sufficient for any one conscious percept?
Focusing on the sense of phenomenological consciousness, research suggests that this experience does not require behavior, emotions, selective attention, language, self-consciousness, or long-term memory.
However, it is dependent on some complex biological system such that destruction of certain cortical regions interfere with the contents of consciousness.
In addition to exploring different areas that have been postulated to be responsible for consciousness (the claustrum is an emerging candidate), this talk discusses critical, and necessary, background conditions that must be met \textit{along with the neural correlates} for a biological organism to function and experience consciousness.
The neuroscience of cognition and consciousness extends beyond humans, and it is possible to see the remarkable similarities across neural structures and architectures between humans and other animals.
Exploring the neural underpinnings of consciousness raises a number of hard questions regarding how and when consciousness emerges in beings, and how it relates to biological and machine intelligence (if at all).

\paragraph{Giulio Tononi:  Integrated Information Theory: from Consciousness to its Physical Substrate}
To be conscious is to have an experience.
One cannot start from matter and ``squeeze'' consciousness out of it, but one can start from consciousness itself and ask what physical system could account for its properties.
Integrated Information Theory (IIT) starts from phenomenology, not from behavioral or neural correlates.
It identifies the essential properties of every experience (axioms), derives the requirements that physical systems must satisfy to account for them (postulates), and has explanatory, predictive, and inferential power.
Phi ($\phi$), as a quantitative measure of integrated causality (consciousness), is high for cerebral cortex, but low for other complex areas such as cerebellum and the basal ganglia due to the nature of the neural connections and architecture.
Translating to practice, it is possible to test for consciousness by using a transcranial magnetic stimulation (TMS) pulse to ``ping'' the brain with external activity, then measure the causal changes in activity as a metric for \textit{cortical effective connectivity} \cite{massimini2005breakdown}.
This technique is being tested in patients \cite{casarotto2016stratification} and is currently named the \textit{Perturbational Complexity Index (PCI)}.

\paragraph{Owen Holland: How Neuroscience, Cognitive Science, and Psychology can provide what Machine Consciousness needs, and how Machine Consciousness can pay them back}
Machine consciousness (MC) is an engineering discipline, and theories from consciousness science can help guide the implementations that these engineers pursue.
Full theories of consciousness should be able to posit that in a system of type W, some components X will manifest characteristics of type Y under some circumstances Z.
While we understand that the human brain is an instance of W, the other required components to complete this theory are open-ended and not understood.
Engineers and theorists have provided some suggestions for X (neural correlates of consciousness) and Y (see, Thomas Metzinger, Igor Aleksander), but as an engineering discipline, MC requires all of these components to be detailed enough to implement in a real system (see LIDA \cite{franklin_lida_2006}).
Even more recent theories of neural function, such as the predictive processing model that suggests the brain is a probabilistic hierarchical generative network, does not provide sufficient detail to guide MC engineers in implementing a system.
Even if these are correct, they do not address the hard problem of phenomenological emergence.
So, where are the MC projects going to be carried out?
Who is a contender?

\subsubsection{Breakout Discussion}
Because of timing and scheduling conflicts, the Breakout discussions took place prior to the final presentation (Owen Holland).
This is noted to reflect that the breakout discussions would not have had content from the final talk to discuss.

\paragraph{Breakout 1}
The distinction between types of consciousness is critical for accurately characterizing consciousness, as it is necessary to align the conversation depending on whether one is discussing phenomenological, access, or human-level consciousness.
There was some debate as to whether access consciousness is clearly distinct from phenomenological consciousness, or if it is merely the contents of consciousness that differ between them.
Phenomenological consciousness may actually have some causal power, as it could be supporting anticipatory contentful flow (the ability to construct a semantic or functional flow to otherwise disparate content).
This form of consciousness is likely dependent on the available phenomenological space (i.e., \textit{umwelt}).
Following this, the mechanisms that support consciousness are potentially the neural correlates that are jointly sufficient for phenomenological consciousness.
Although useful for AI, it was generally agreed upon (not universally) that the hardware-software distinction is not helpful for understanding human consciousness as we know it.
As a potential consequence of neural correlates, one can argue for an \textit{intermediate level theory} of consciousness.
This theory posits that phenomenological consciousness resides between low-level processing and higher-level abstraction and humans are experiencing the intermediate representation that is influenced by processing of perceptual and abstract concepts.
Work is needed to understand how, and if, this is in agreement with higher order thought and human-level consciousness.
The group found the Perturbation Complexity Index and $\phi$ to be promising measures of consciousness, even if they were not absolute.
Additionally, it was argued that the creative ability to discover and prove answers to certain hard problems may imply phenomenological consciousness.
While machine consciousness may be possible in the future, and this raises questions about ethics, rights, and testing for consciousness, the group agreed that machine intelligence is a concern that should be monitored closely.

\paragraph{Breakout 2}
Characterizing consciousness typically begins with the assumption that a ``self'' or ``I'' (e.g., Descartes' Cogito) is required.
In other words, start with consciousness, and infer rules from there.
Towards this, the group discussed breaking apart the notions of selfhood and phenomenological consciousness such that one might just argue that experience exists.
Therefore, something that has experiences exists.
In fact, it may be that the self is just what we label conscious experiences connected through time, suggesting that self-consciousness is a function of memory.
The underpinnings of consciousness may be the robust neural ``chains''  that build to grid structures in a $\phi$-like system.
In this sense, the topology of the system may change the flavor of consciousness.
Beyond structure, it may be that time (or flow) is a fundamental axiom of consciousness.
Following the mechanistic underpinnings, the group agreed that $\phi$ or $\phi_E$ were reasonable potential tests for consciousness.
Machine consciousness may be able to exist in principle, but now right now.
Even if we simulate the whole brain, we are unlikely to capture causal connections necessary for consciousness.
We want the thing that exists for itself, not just what seems to extrinsically exist.
Machine consciousness, if functional and not phenomenological, could lead to useful labor without guilt.
However, care should be taken not to over-attribute consciousness to a system that may not have it, as this may lead to an incorrect prioritization of equipment over human lives.

\subsection{Focused Session 4: Computation and Logic}
\label{sec:fs4}
The fourth focused session discussed computational, mathematical, and physics-based formalisms.
This session explored different theories of consciousness, particularly focused around computation and formal mathematics.
Formalized axioms to support consciousness (at least human-level cognition) were presented and discussed, along with explorations of information integration and how, exactly, to define AI as a discipline.
Focused Session 4 featured a total of eight presentations.
The external workshop attendees included;

\noindent\textbf{External attendees:} Henk Barendregt, Susan Blackmore, Selmer Bringsjord, Antonio Chella, David Gamez, Owen Holland, Subhash Kak

\noindent\textbf{SRI attendees:} David Israel, John Murray, David Sahner, Natarajan Shankar, Damien Williams

\subsubsection{Presentations}
\paragraph{David Gamez: Informational, Computational and Physical Theories of Consciousness}
Theories of consciousness have evolved over time with modern theories being focused around the physical world.
Consciousness, being only reportable through self-report, is problematic from a measurement standpoint.
Because we can accurately measure the physical world, we need appropriate models that allow us to translate physical world measurements (p) to the world of consciousness (c).
An appropriate model, or \textit{c-theory}, is a mathematical description of the relationship between measurements of consciousness and measurements of the physical world.
To date, Information Integration Theory (IIT) is the closest thing we have to a c-theory.
While one can propose information, computation, or physical theories of consciousness, here it is argued that information and computation fall short because they can be re-interpreted through physical theories, suggesting that only physical theories are viable.
If one were to use computational methods to determine and develop c-theories, what would be the best strategies and mathematics for doing so?
Is it possible that these theories can describe biological structures in a way that enables generalization to future artificial systems?

\paragraph{Henk Barendregt: Axiomatizing consciousness with applications}
The explanatory gap between our ability to model consciousness from a third person perspective and the true first person perspective of consciousness is dubbed the hard problem, and it will not be solved during this talk.
Rather, this talk provides axioms that model aspects of consciousness, motivated by neuropsychology, computer science, and Vipassana meditation.
In particular, fallacies inherent in consciousness, and how its intimate relationship with time affects the human sense of free will.
Consciousness is not as it seems.
It is discrete, despite feeling continuous.
It feels veridical, but is often incorrect.
Humans perceive information that biases behavior outside of awareness.
The illusory sense of continuity necessitates interpolation of information and selective omission of information, and this process can explain why consciousness is not veridical and why omitted information is not experienced despite being processed.
This demonstrates that time is a critical factor, as consciousness depends on time.
Consciousness has an object, an intended action, and a state that maps the object to the action.
Our actions change the world state, and consciousness may be viewed as the predicted configuration stream of the world that based on all of these inputs (world, state, object, action), which is continuously being updated and corrected.
If one considers strange attractors that promote repeated and predictable world and configuration states, mindfulness is the process of deconditioning this circular state of existence.
This has bearing on the definition of free will, as the attractors and cravings humans experience lead to what should be called \textit{restricted} free will.

\paragraph{Selmer Bringsjord: Axiomatizing consciousness via eleven formulae \& linking to axiomatic physics}
\textit{Author-Supplied Abstract:} ``We present the axiom system CA, intended to formalize human-level cognitive (HLC) consciousness and associated mental phenomena.
Like our work in axiomatic physics, in which computing machines can be enlisted to help discover and confirm illuminating theorems, CA is intended to be implemented.
And, one of the axioms of CA forges a direct connection between cognition and axiomatic physics, since it asserts that human agents understand causation as it is captured in some axiomatization (possibly na\"{i}ve) of part of physics.''
The eleven consciousness axioms, elaborated in \cite{bringsjord_toward_2018}, are;
\textit{(1)} Planning (following \cite{aleksander_axioms_2003}),
\textit{(2)} Perception-to-Belief (P2B; perceiving is believing),
\textit{(3)} Knowledge-to-Belief (K2B),
\textit{(4)} Introspection (if you know something, then you know that you know it),
\textit{(5)} Incorrigibilism (identifying as having a property because you believe it so),
\textit{(6)} Essence (unique sense of self),
\textit{(7)} Non-Compositionality of Emotions (an agent can enter an emotional state, but these are not all constituted by some conjunction of ``building-block'' emotions);
\textit{(8)} Irreversibility (conscious processing is irreversible),
\textit{(9)} Freedom (agents perceive or believe they are free),
\textit{(10)} Causation, and
\textit{(11)} TheI (self-belief).

\paragraph{David Sahner: Contemporary Physics and its Nexus with Consciousness:  Should a Theory Wait? }
The predictive powers of contemporary physics are no doubt impressive, and quantum theory is potentially one of most impressive theories of all time.
However, as successful and powerful as quantum theory is, discussions of a quantum theoretical basis for human consciousness are likely to be premature.
In biological systems, quantum theory has been evoked to explain things from photosynthesis to avian navigation.
Moving to the realm of human consciousness, one can place consciousness in the position of the observational system that interacts with the quantum world.
However, given the conflicts and inability to reconcile general relativity and quantum theory, is it reasonable to even think of introducing human consciousness into the conversation?
The back half of this talk focuses on modern theories of quantum mechanics, and how science may be working towards agreeable theories.

\paragraph{David Israel: Some Thoughts on Goals and Methodologies in Artificial Intelligence}
Just what kind of discipline is \textit{Artificial Intelligence (AI)}?
If it purely mathematical, then do we focus on, and judge the validity of, proofs and rank the quality of theorems to determine impact?
If it is an empirical science, then does it generate falsifiable hypotheses that are (dis)confirmed through experimentation?
Since the seminal NSF-Funded Dartmouth Workshop debate in 1956, the field has been shaped by disagreements in what AI should be.
Simon \& Newell emphasized cognitive science, focusing on abstract human problem-solving (note, this ignores lower level perception and motor skill functions).
McCarthy promoted the highly influential approach based on logical AI; a focus which stressed math and has had little impact on cognitive psychology.
This talk makes the case that AI is a \textit{design} discipline; one where systems are motivated by (but not locked to) cognitive psychology in order to get systems to perform complex processes intelligently.
This goal, of developing systems that can do \textit{some} things intelligently (not all!), says little about consciousness, but stresses the cross-disciplinary nature of artificial intelligence.

\paragraph{Subhash Kak: Quantum Mechanics and the Consciousness Problem}
If the scientific method relies on reductionism, is consciousness, which is irreducible, amenable to scientific study?
This talk links the problem of measuring consciousness to quantum mechanics and the complexities of measuring systems that behave in ways that are not intuitive.
Observation is said to collapse the wave function in quantum mechanisms, so is it possible that consciousness can act as the observer and become a fundamental part of physics?
This discussion covers a breadth of topics in quantum mechanics, from Schr\"{o}dinger's equation and thought experiments to polarized waves and beam displacement.
The talk ends with a discussion of future ideas and directions in the realm of quantum research, from computing to teleportation and AI.

\paragraph{Susan Blackmore: The Big Question for Consciousness
Studies} Is consciousness something added to all the functions and
abilities we (and maybe other animals and machines) have, or is it
intrinsic to those functions and abilities?  This question is of real
significance because it concerns the basis of phenomenological
consciousness; its function and evolution.  Many theories of
consciousness, such as Global Workspace Theory and Information
Integration Theory, are of the opinion that consciousness is
``something extra'' that emerges from some aspect of brain function
but goes beyond the sum of the parts.  
Other
theories posit that this is not the case, such as Illusionism and
Eliminative Materialism.  These competing views are discussed, with a
framing around Daniel Dennett's Multiple Drafts Model.  Depending on
one's take, this has implications for machine consciousness.  Machines
will be conscious like us only when they are as deluded as us.

\paragraph{Natarajan Shankar: Explorations in Robot Consciousness From Integrated Information to Information Integration}
Science is what we understand well enough to explain to a computer.
Art is everything else we do.
Akin to this dichotomy, consciousness has solvable and unsolvable problems.
Phenomenological consciousness, subjective experience, is hard to study objectively and without a testable definition, we do not know that machines lack it.
Access consciousness, on the other hand, is about turning inputs into information, which machines clearly can do.
Taking all of this together, humans are amazingly capable of taking inputs and integrating them in ways that keep them alive to continue experiencing the world.
Relating consciousness to intelligent agents, consciousness as an integrated experience allows an agent to react in a unitary way to error/anomaly/reward, i.e., to take responsibility.
This talk presents the concept of the Integrated Information Theory of Consciousness (IITC), defining consciousness as the level of integration across modules.
By borrowing from information theory and cognitive psychology, the theory of \textit{Cueology} is presented, which explains how we can create ``meaning machines'' through cue-based robotics and systems.

\subsubsection{Breakout Discussion}

\paragraph{Breakout 1}
Consciousness can be characterized with two types.
Little C: Cognitive consciousness; the ability to engage in or emulate useful complex behavior.
This is robust, functional cognition and is likely to track with intelligence.
Big C: Phenomenological; the majority of the discussants thought there was a Big C.
Towards the mechanistic underpinnings of consciousness, the group believed that the theories should be falsifiable, and many theories have a number of drawbacks (from quantum to multiple drafts).
Measuring consciousness may be possible through neural correlates, but the group conceded that no single correlate is likely to be found to support Big C, particularly given the current technological limits.
To extend on this, one could assess behavioral and social correlates.
However, many agreed that Big C could not reliably be probed ($\phi$ for instance, is not adequate).
They also agreed that Machine Little C is achievable in the future, but there was disagreement as to whether Big C would ever be achievable.
Most agreed that autonomous, multi-task-capable systems pose danger, but through careful formal verification it is possible to mitigate negative outcomes.

\paragraph{Breakout 2}
This group had a broad discussion on what Big C (phenomenological) might be, and how it could be probed.
In defining consciousness, it is worth considering the embodiment arguments as well, such that experience has strong bases in sensorimotor theories.
Of note, despite this embodied framing, it does not necessarily limit consciousness such that many senses are ignored, and out-of-body-experiences as well as culture may affect how we frame embodiment.
There are multiple potential ways of probing Big C, such as meditation or assessments of subjective estimates of time and the environment.
From a machine consciousness perspective, it raises questions as to how a machine with Big C (if possible) could ever explain, define, or describe its own consciousness.
Towards affording machine consciousness rights, it is argued that ethics is rooted in sentience, but that we, as humans, do not have the cleanest track record (e.g., racism/bigotry).

\subsection{Focused Session 5: First-Person and Non-Western Perspectives}
\label{sec:fs5}
The fifth focused session covered first-person and non-western philosophical perspectives.
This session explored the qualities that define and modulate phenomenological consciousness.
From a human perspective, this includes altered states of consciousness (whether through pharmaceutical means or intense life events) and the training and focus on consciousness to separate conscious experience from the contents of consciousness.
For the future of machine AI, it was demonstrated how a number of qualities typically assigned to consciousness in the realm of self-reference can be actualized through formal logic, and the question was raised of whether it was possible to imbue a machine with a sense of contentment and mental quietude.
It is possible to link concepts of machine processing and human processing in the sense that both systems and human brains are developed for a particular type of processing and subsequent output.
Mindfulness is the effortful attempt to limit ``automatic'' processing in order to be fully sentient, aware, and conscious of behaviors and mental processes.
Therefore, would a machine AI require a similar capability for mental quietude in order to achieve consciousness?
Focused Session 5 featured a total of nine presentations.
The external workshop attendees included;

\noindent\textbf{External attendees:} Naveen Sundar Govindarajulu, Marcia Grabowecky, Susan Kaiser-Greenland, David Presti

\noindent\textbf{SRI attendees:} Patrick Lincoln, John Murray, John Rushby, David Sahner, Daniel Sanchez, Damien Williams, Adrian Willoughby

\subsubsection{Presentations}
\paragraph{Naveen Sundar Govindarajulu: On Formalizing Eastern Perspectives}
This talk presents a ``self-aware'' agent, Cogito, using formalized cognitive consciousness, and provides three contributions.
Contribution 1: Showed a real-world capability (ethics) that needs self-awareness.
Contribution 2: Showed that self-awareness is essential for building moral machines (as posited in Vedic traditions).
An implementation and formal framework for achieving such machines.
Contribution 3: Using key insights from a formal model of Advaita, a nascent formal system building upon prior * operator.
Through this framework, an extension on previous instantiations of the doctrine of double effect is shown, demonstrating the criticality of incorporating self-reference to ensure formal consistency.

\paragraph{David Presti: A radically-empirical perspective on mind-matter relationship}
If matter is the physical stuff (mass, energy, particles) and consciousness is mental phenomena, what are the ways forward for the scientific study of consciousness?
First, continued probing of the structural and functional aspects of brain and body.
This is a relevant endeavor because you are unlikely to do better or out-think evolution; if we thought of it, probably so did evolution.
Second, interface with fundamental physics.
Third, refined analyses of mental experience.
This can be achieved through studies of mental control, such as researching meditation and Buddhism.
Additionally, scientists should explore altered states of consciousness, potentially leveraging psychedelics.
Lastly, Radical empiricism.
Radical empiricism is the study of less traditional subjects, such as psychical research and non-physical phenomena (near-death experience, out of body experience).
A brief overview of psychical research is provided, with an emphasis on near-death experiences and ``spiritual'' experience.
These events are reported to have prolonged aftereffects, such as: a decreased fear of death, an increased appreciation for life, increased compassion, altruism, and interpersonal connection, and a decrease in materialistic attachment and competitiveness.
Exploring radical empiricism may hold the key to the next big paradigm shift in the scientific understanding of consciousness.
A lively discussion followed the talk, discussing how to ethically study these transcendent, spiritual experiences.
Some examples were given, such as: experiencing the birth of a first child, psilocybin studies \cite[for example]{griffiths2008mystical}, or transient neural stimulation to increase feelings of ``oneness'' with the world \cite[following the ``God Helmet'']{tinoca2014magnetic}.

\paragraph{Marcia Grabowecky: Approaches to Direct Experience of the Mind}
The ability to study consciousness and subjective experience can be guided by Buddhist practices that focus on mental development through meditation.
The long eastern religious traditions of careful subjective analysis of consciousness can provide both hypotheses and data for scientific investigations.
Attention training practices from meditative traditions appear to:
1) be ways to train attention that generalize,
2) potentially stabilize mental states by reducing neural noise while maintaining alertness, and
3) be used to explore the nature of conscious awareness during stages of sleep that are usually unavailable to subjective report.
This talk provides an overview of meditative practices, both concentrative focused attention and vigilant mindfulness, and shows how seemingly automatic perceptual processes can be altered based on empirical studies of long-term practitioners.
Additionally, an overview of lucid dreaming is discussed, providing insight into how consciousness and subjective experience manifest without direct external input.

\paragraph{John Rushby: HOT Theories of Consciousness, And Speculation on Evolutionary Origins}
\textit{Author-Supplied Abstract:}
``HOT theories of consciousness associate consciousness with ``thoughts about thoughts.''
These are referred to as ``Higher Order Thoughts,'' hence the acronym HOT.
I will sketch the basic idea of HOT theories, and outline some of its many different variants.

I then speculate on the evolutionary origin of consciousness, and its purpose.
A unique attribute of humans, and the reason we dominate the world, is our capacity of teamwork; this requires ``shared intentionality,'' the ability of one human to create a new goal and communicate it to others.
The new goal begins as a (likely unconscious) cluster of mental activity; to communicate it to others, it must be abstracted to a succinct form\ldots e.g., concepts.
I argue that this leads naturally to a dual process model of cognition where the ``upper'' level comprises (abstracted) thoughts about (unconscious) lower level thoughts.
As in other HOT theories, this model associates consciousness with the upper level thoughts.''

\paragraph{John Murray: The Fringes of Consciousness}
Many of our perceptions and behaviors do not occur consciously, but just on the fringes of consciousness.
Despite our best conscious efforts, sometimes we cannot help but blush, and sometimes we have information just on the tip-of-our-tongue that \textit{we know we know} but cannot recall.
These examples suggest that consciousness is not necessarily boolean; there appears to be a continuum.
Distortions of consciousness, through drug-induced spatiotemporal hallucinations, further demonstrate that experience is not fixed.
Towards ensuring a robust and safe AI future, it is necessary to consider that consciousness is not something that operates in a specific and pre-defined way.
Rather, it manifests in complex ways we cannot always anticipate, motivating the need for appropriate oversight and governance as we build future intelligent systems.

\paragraph{Susan Kaiser-Greenland: The Headless Path: How to Build an Enlightened Robot}
From a self-professed ``recovering lawyer,'' this talk discusses how meditation and mindfulness can be used for 1) stress reduction and self improvement, and 2) liberation and psychological freedom.
The headless path is about discovering the lack of separation between oneself and the rest of the universe, in that nothing exists in isolation.
Through wisdom and compassion, meditation focuses one to understand objective truth and to separate the judgmentality and emotions that we typically associate with our experiences as being ``add-ons'' that we unwillingly attach, which can be separated and processed separately.
By doing so, an agent, or individual, is able to get to ultimate truth and separate the reflection from the truth itself.
During discussion, the headless path was compared to Kabat-Zinn-style ``non-judgmentality'' such that feelings and emotions (judgment) are not meant to be put away, but are meant to be put aside and processed separately, in order to deeply understand why they are associated with events and things in the world.
Additionally, comparisons are made between the motivating mentality that everyone is inherently good, compared to the Judeo-Christian view of Origin Sin.

\paragraph{Damien Williams: A Discussion on Taoism and Machine Consciousness}
This talk provides an overview of Toaism (\textit{The Way}) and presents philosophical paradoxes and thought puzzles that are prominent and influential from eastern philosophy and scripture.
Emphasizing principles and writings from Chuang Tzu and Lao Tzu, one can understand the cultivation of consciousness as a return to simplicity and the ability to accept the importance of action without action, or \textit{Wu-Wei}.
Wu-Wei is the powerful concept of understanding when not to act, as this provides introspection and insight into the natural desire to move or act towards a goal, which can sometimes be harmful and push one farther away.
Towards a future AI, it is worth considering how such principles could be incorporated into machine consciousness.
Could a machine understand contentment and pursue Wu-Wei?

\paragraph{Daniel Sanchez: Revealing the Misconceptions of Perception}
Buddhist principles of the path teach that the mind can be trained through meditative practices of concentration and open awareness.
Through this training, it is possible to realize the lack of self and to understand that you, as the perceiver, are not independent or separate from the stimulus and environment you experience.
This lack of consistent ``self'' is a difficult concept to understand, in that we, as humans, typically feel like we are independent, unique, and perceive reality as we understand it.
However, illusions (optical or otherwise) are clear demonstrations that humans do not perceive the environment in a veridical fashion, and that these mis-perceptions are typically shared by others, demonstrating a shared perception or experience.
Following the concept that illusions are shared human perceptions, this shows the similarity of human conscious experience due to our shared perceptual machinery.
In contrast, adversarial networks show how fundamentally different machine illusions can be, in that the misclassifications that machines make are not intuitively similar to human illusory constructs.
Given this disparity, it can be reasoned that machine consciousness, if ever realized, will be fundamentally different from humans because of the vast gap between our perceptions of the environment.
Towards understanding machine consciousness, it is argued that machine vision systems should be developed to both classify and misclassify as humans do in order to move towards a more human-like future consciousness.
In other words, human optical illusions are a reasonable Turing Test for visual AI.

\paragraph{Christopher Connolly (via Daniel Sanchez): Basal Ganglia}
This presentation was put together by Christopher Connolly, but due to scheduling conflicts, was delivered by Daniel Sanchez.
Daniel had no prior knowledge of the content or narrative, but volunteered to lead a discussion on the content as his expertise was best aligned with the domain.
The basal ganglia (also known as the striatum) is a phylogenetically-old memory system that provides a reward-based learning mechanism based on looping connections with the cortex.
The basal ganglia has tremendous integrative capabilities; 30,0000 cortical neurons can converge on a single striatal neuron.
Despite being a neural region not necessarily considered a ``neural correlate of consciousness,'' disorders of basal ganglia function have been shown to produce alterations of conscious perception of space and time.
Parkinson's Disease, a pathology based on dysfunction of the neurotransmitter, Dopamine, due to cell death in the substantia nigra, leads to behavioral impairments that participants are \textit{consciously aware of}.
Huntington's Disease, a pathology based on dysfunction with striatal neurons, leads to motor impairments and cognitive disruptions (e.g., disorganized thought).
While these diseases are notable and recognized for a variety of impairments, these effects on conscious processing and subjective experience are not typically highlighted, and demonstrate how disrupted processing of inputs and information lead to significant disruptions in conscious experience.

\subsubsection{Breakout Discussion}

\paragraph{Breakout 1}
This group presented the idea of the \textit{Philosophical Snowman}.
The largest, bottom sphere represents the inputs and representations while the middle sphere represents models of information and the world, including the self.
The top, smallest sphere, or head, is where pure consciousness resides and interacts with models of the world.
The lack of a compelling argument for the underpinnings of consciousness opens the door to speculation about consciousness fields and other mechanisms that we do not well understand, such as gut biomes.
Measuring consciousness then, was equally broad.
Once we understand consciousness in the future, we can test it by perturbing its components, whether it be a consciousness field or some underlying biological construct.
The potential for machine consciousness, given the ambiguity about what constitutes and leads to consciousness, is there, if we can ever discover it.
If ever realized, it could provide a huge service to mankind and the planet -- particularly if imbued with Wu-Wei.

\paragraph{Breakout 2}
Towards understanding phenomenological consciousness, this group addressed the question of how one could become aware of being sentient.
Given the amount of time humans spend ``doing,'' it is very easy to automate complex behaviors and separate conscious experience from the activities of everyday.
This is, essentially, the situation that mindfulness attempts to address.
As our thoughts and mental processes become overly concerned with higher level cognition (planning, reasoning), our body is operating in a procedural mode such that the contents of our consciousness are being driven by needs and requirements, and are not volitionally controlled.
Meditation is one way to become aware of phenomenological consciousness, but it could alternatively be cued through sensations such as hypnogogic jerk or an extreme external input (trauma or physical input).
Another way is through shared reporting of consciousness and experience; we have a presupposition that the physical world is real, so by being willing to recognize that we never experience the world directly, we learn by shared reports.
This type of consciousness, then, is defined by awareness of the present moment; whether it is external or internal (thoughts) experience.
It is critical to note this is separate from the \textit{contents} of consciousness.
Because of the separation of phenomenological consciousness from other cognitive and physiological activities, it may be that no true measure of consciousness exists.
Towards this, because we do not understand how consciousness emerges, it is not clear how any machine or simulation of consciousness would acquire the actual property of consciousness.

\subsection{Focused Session 6: Machine Consciousness}
\label{sec:fs6}
The sixth focused session was on Artificial Intelligence and machine consciousness.
This session explored the characterization of advanced machine intelligence, distinguishing between autonomy, intelligence, sentience, and consciousness (cognitive and phenomenological).
Additionally, theories of ethics, morality, and rights that should be afforded to these systems was discussed, with care given to take these various aspects and characterizations into account.
For instance, what moral obligation do humans have to an intelligent system versus the same system with (or without) sentience?
Focused Session 6 featured a total of nine presentations.
The external workshop attendees included;

\noindent\textbf{External attendees:} Selmer Bringsjord, Antonio Chella, David Gamez, Owen Holland, Jonathan Moreno, Ron Rensink, Susan Schneider, John Sullins, Shannon Vallor, Robin Zebrowski

\noindent\textbf{SRI attendees:} Boone Adkins, David Israel, Kellie Keifer, Patrick Lincoln, John Murray, Karen Myers, Andy Poggio, John Rushby, David Sahner, Damien Williams

\subsubsection{Presentations}
\paragraph{David Gamez: From Human to Machine Consciousness}
Machine consciousness (MC) can be delineated based on varying levels or categories (note, the levels are not exclusive).
MC1 are machines with the same external behavior as conscious systems; akin to Chalmer's \textit{Philosophical Zombie}.
Many systems have been created that behaviorally replicate seemingly conscious behaviors, such as IBM's Watson, however they make no claim to have internal consciousness.
MC2 are computer models of the correlates of consciousness.
Through robotic and simulation systems, such as CRONOS (and SIMNOS; Owen Holland) and NeuroBot, large spiking neural networks were created to simulate biological eye control and global workspace concepts.
However, while these systems tell us about physical systems, they do not actually speak to machine consciousness at any level.
MC3 are computer models of consciousness (models of bubbles of experience).
Systems, such as those supporting Sony's AIBO dog robot and Antonio Chella's Cicerobot, have specifically been designed to mimic a conscious agent's abilities to represent the environment and their interactions within it.
However, while these systems model behaviors and representations typically associated with consciousness, they do nothing towards creating actual consciousness.
Lastly, MC4 are machines that are actually conscious.
We have yet to develop machines we might argue are conscious, so we use human consciousness as the standard for assessment.
C-Theories are proposed; models that translate the description of the physical state into formal descriptions of the conscious state.
These theories can be developed based on the human standard, and applied to subsequent machine systems and future AI to assess predicted machine consciousness.
From an ethical standpoint, MC4 systems may necessitate rights and ethical treatment, but MC1 machines should be treated carefully as they could threaten humanity with their combination of powerful behaviors and lack of consciousness.

\paragraph{Antonio Chella: A Cognitive Architecture for Robot Self-Consciousness}
Towards a self-conscious robot, an architecture is presented that consists of (1) sensory data inputs to a (2) sub-conceptual area that feeds into a (3) conceptual area and, finally, a (4) linguistic area.
The architecture is a hybrid, combining a formal knowledge base with connectionist neural networks.
Focus is given to how conceptual representations are built, stored, accessed, and used.
A hammer is used as an example of how the visual input of simple geometry and relational structure gives way to functionality and conceptual representation at a higher level.
This architectural framework is used as the basis for higher-level concepts including theory-of-mind and self-consciousness.
A working example based on this architecture is shown, with Cicerobot.
Future robotic work is shown, using the Pepper robot as the basis for a future self-conscious machine.

\paragraph{Ronald Rensink: Consciousness, Attention, and Machines}
The question of whether a machine could be conscious, and how we would know, is too difficult to tackle head on, so here it is proposed to focus on \textit{fractionation} -- the functional examination of the atomic components of visual experience, or primary consciousness.
This approach follows from David Marr's framework of analyzing an information processing system based on its 1) function, 2) representation, and 3) implementation.
With further fractionation, within the specific domain of visual experience, this talk focuses in even further on the concept of visual attention.
Attention, like visual experience, can be considered a complex \textit{adjective}, rather than a noun, and can be divided into numerous components based on spatio-temporal features and varying task-relevant functions.
Using a taxonomy of visual attention applied to consciousness, it is possible to relate different kinds of visual attention to different layers of visual experience.
For example, attentional binding relates to assembled/static image perception while attentional holding relates to coherent, continuous visual experience.
What follows is the \textit{Necessity thesis}: an attentional process of type \textit{m} is needed for conscious experience of type \textit{n}; each type of conscious visual experience enables the control of its associated selective process, based on global considerations.
It is argued that this thesis can be applied to other types of consciousness; whereby other forms, such as subjective or introspective consciousness, can similarly be fractionated and operationalized for subsequent study.

\paragraph{Robin Zebrowski: Extended Mind Metaphysics and the New Challenge for Artificial Intelligence}
Is it possible to separate out the contributions of the individual from the contributions of the environment when discussing cognition and consciousness?
Here, an argument is presented that describes the importance, and potential necessity, of considering embodiment when attempting to understand and model consciousness.
It is argued whether the embodiment argument ties to, or alters, debates around functionalism, but it raises a critical point for AI and machine consciousness.
In particular, metaphysics of the mind matters for machine consciousness, so we cannot ignore the hard questions, and must consider how the ``extended mind,'' or environment, impacts the representations of an agent.
One could argue that ``at each step we should build complete intelligent systems that we let loose in the real world with real sensing and real action. Anything less provides a candidate with which we can delude ourselves'' \cite{brooks_intelligence_1991}.
This talk explores the critical interaction between the agent and the environment in which it inhabits.

\paragraph{Susan Schneider: Tests for AI Consciousness}
Testing for phenomenological consciousness in a machine is a, if not \textit{the}, hard problem.
Towards an answer, the AI Consciousness Test (ACT) is proposed \cite{schneider-is-2017}.
ACT is a question-and-answer test that is based on the premise that adults can \textit{quickly} and \textit{readily} grasp concepts based on the quality of experiential consciousness.
ACT would challenge a potentially conscious AI with increasingly complex questions to determine if the system could conceive of hard problems, such as self-hood, morality, or consciousness itself.
While other tests exist, such as using $\phi$ from Integrated Information Theory to measure the potential consciousness of a silicon chip, we do not yet know how machine consciousness will manifest, so it reasons that we should use tests similar to those we would use on a functioning human today.
It is argued that this test is relevant for two components of consciousness engineering; optimal AI performance and ethical use.

\paragraph{Jonathan Moreno: Is Mind Expansion a Test of AI Consciousness?}
While it may not be a necessary condition for AI consciousness, might the capacity for mind expansion be a test for conscious AI?
Psychedelics are a canonical example of a method that humans use to expand their phenomenological experience.
Thus, understanding how and why humans use psychedelics, and the impacts they have, may help guide our understanding of consciousness.
This talk covers a history LSD and its potential uses, including the exploration of the psyche by ``psychonauts'' and their further-reaching compatriots, the ``cyberpunks.''
Nascent research on LSD shows changes in neural activity that corresponds to changes in experience and phenomenological consciousness.
This research may be a new venue for understanding how the brain gives rise to consciousness, and the desire and ability for systems to expand their consciousness may provide value in defining a fundamental drive associated with conscious systems.

\paragraph{Selmer Bringsjord: Extremely Emotional Robots}
\textit{Author-Supplied Abstract:} ``We reveal real robots with extreme self-knowledge, including deep and far-reaching knowledge of the limits of their knowledge.
We anchor this engineering by considering Implemented versions of Alonzo Church’s Knowability Paradox, which purports to shockingly deduce that from the solitude \& comfort of a humanoid robot’s armchair, it can deduce that there are truths that are mathematically impossible for it to know.''

\paragraph{David Israel: Robo-Ethics: A View from AI}
It is argued that robot ethics is not needed, and the ethical quandaries that arise due to intelligent robots and AI are not unique.
Rather, the focus should be on issues related to the moral agency and responsibility of increasingly autonomous artifacts, paralleling situations such as the creation of the atomic bomb or offspring (children).
What is critical to examine is how artificial \textit{intelligence} should behave, in the sense that it has the ability to act intelligently, and, therefore, the responsibility to act morally.
Thus, consciousness is not necessarily the issue, so much as the ability for an autonomous agent to make decisions.
This decision making ability is framed based on agent preferences, goals, and, utility.
A discussion followed on the relevance of addressing the ethics of autonomous and intelligent AI towards providing funding agencies (such as DARPA who dealt with ethical issues related to neuroengineering the early 2000's \cite{noauthor_silence_2003,rudolph_military_2003,rizzuto_militaryfunded_2003}) with guidance on research and development prior to problems emerging.
Additionally, comparing intelligence, but non-conscious, systems to children (who are arguably conscious) was questioned by some.

\subsubsection{Breakout Discussion}

\paragraph{Breakout 1}
Consciousness is best characterized as a coordination of types.
Whether described as a ``cluster concept'' or a fractionated whole, consciousness can be thought of as a ``coordinated community of components'' rather than a singular or integrated entity.
The reason for dissociated integration from coordination is that consciousness seems to be plastic, or malleable, and is an ever-changing construct based on physical, social, and cultural factors that can affect it in important ways on various time scales (identity can change over long and/or short periods of time).
Because consciousness is the result of dynamic processes, time is a critical component that must be taken into account as a mechanism of consciousness.
The experience, compression, dilation, and conception of time impact consciousness on a fundamental level, and these factors are, unfortunately, not well captured by formalisms or logic.
Because of the many types of consciousness, there must be many tests to assess it properly.
From $\phi$ to introspection and the experience of humor, it may require a varied battery of tests to assess a conscious agent.
One component that is worth exploring, is how time, which is argued to be critical for consciousness, can be used to assess conscious experience.
This all results in a complex answer regarding the realization of machine consciousness.
It may be possible, but it is critical to discuss consciousness as its fractionated components (visual perception, memory, attention, etc) that may be realized, rather than expecting a singular type of consciousness such as the one we are used to in humans.
Humanity's relationship with any sort of machine consciousness will have to be considered carefully, as even human-to-human communication can be misinterpreted despite a shared type of consciousness.
The group did raise questions about how we should approach a potential machine consciousness (independent of type); should it be given full autonomy, a sense of purpose or morality?
Or should these systems be allowed to develop these senses on their own?

\paragraph{Breakout 2}
Although interdisciplinary approaches for characterizing consciousness are useful, it is argued that not all disciplines are needed.
There was not agreement on approach, but one side argued for empirical science with a reduced emphasis on philosophy, while others argued that even if you remove philosophers from the discussion, philosophy will remain a part of the conversation.
Towards empiricism, the group stressed a focus on cognitive consciousness rather than phenomenological.
Phenomenological consciousness may not have mechanistic underpinnings or an actual function, whereas cognitive consciousness -- or what is \textit{important} about consciousness -- may be thought of as a collection of interacting cognitive systems.
Given the focus on cognition, the group stressed tests for cognitive consciousness, with the argument that tests for phenomenological experience may be impossible.
The group generally believe that machine consciousness is not possible, but that extensive research should be committed to understanding the ethical impact of autonomous systems.

\paragraph{Breakout 3}
This group generally acknowledged that consciousness should be characterized based on the various ``axes'' that exist, following the fractionation argument that consciousness is simply a unitary capability, but a coordination of processes.
The members did not converge on an agreed upon mechanistic underpinning, but considered the various theories that exist.
In general, the group was not sympathetic to illusionism, computationalism, substance dualism, or property dualism.
Towards testing consciousness, the group devised the \textit{Squirrel Test}.
The Squirrel Test is the ability for an agent to produce a self-image of being something other than itself.
In other words, can an agent report on their ability to imagine what it is like to be something else, like a squirrel?
This group generally agreed that machine consciousness was possible; in particular, cognitive consciousness.
However, there was hesitation around whether phenomenological consciousness or sentience was possible, particularly at levels that would rival that of a human.
Independent of consciousness, a machine intelligence with autonomy and power is a potential threat, and it was arguable whether sentience would materially sway the risk-benefit ratio in any particular direction.

\subsection{Plenary Session 2: Summaries, Synthesis, and Research Ideas}
\label{sec:ps2}
This final plenary session focused on summarizing and synthesizing the various topics that were covered during the focused sessions.
This included summary presentations across a number of fields, representing the inter-disciplinary nature of the workshop, and presentations of future project ideas from a number of the attendees.
Plenary Session 2 featured a total of ten full presentations, and 11 ``micro'' presentations that focused on future research ideas.
Considering this was the final plenary, there was a large attendance.
The external workshop attendees included;

\noindent\textbf{External attendees:} Henk Barendregt, Mark Bickhard, Joanna Bryson, Antonio Chella, David Gamez, Naveen Sundar Govindarajulu, Marcia Grabowecky, Owen Holland, Ian Horswill, Subhash Kak, Christof Koch, Jonathan Moreno, Julia Mossbridge, John Sullins, Shannon Vallor, Robin Zebrowski

\noindent\textbf{SRI attendees:} David Israel, Kellie Kiefer, Gordon Kirkwood, Patrick Lincoln, John Murray, John Rushby, David Sahner, Damien Williams

\subsubsection{Presentations}
\paragraph{Subhash Kak: Old Ways of Looking at Consciousness: Current Relevance}
After a brief history of consciousness research, this talk compares and contrasts ``Big-C'' from ``Little-C.''
Big-C is consciousness as an instrument and witness, and its connection to quantum mechanics is presented.
Little-C is consciousness as a process and emergent characteristic, and is within the realm of a strong AI.
These two distinctions are also presented in context of Eastern traditions and philosophies that support differing views on the origins and functions of consciousness (Big-C, Vedanta; Little-C, Buddhism).
This talk covers a variety of concepts that link quantum theories to consciousness, such as the many-worlds-intepretation, decoherence, and the quantum zeno effect.

\paragraph{Damien Williams: Science, Ethics, Epistemology, and Society: Gains for All via New Kinds of Minds}
This presentation provides a broad overview of topics and concepts that were covered and discussed during the six focused sessions during the SRI Technology and Consciousness Workshop.
After this overview, key components related to self and personhood are discussed.
In particular, the criticality of understanding and appreciating the minds, identities, and beliefs of those different than oneself.
By doing so, this may help bridge the hard problem and explanatory gaps, in that it allows for a deep understanding of commonalities and variances, and the potential underpinnings of both.
Per the author, ``If we ever want to create a robustly conscious machine, we should first listen to people who are different from us and who have been systemically prevented from speaking to us, because they know things that we don't.''

\paragraph{Jonathan Moreno: Autonomous Systems and National Security}
How will autonomous systems be involved in future military operations?
Historically, there has been a precedent such that autonomous systems can assist in decision-making and action, but humans should critically remain in (or on) ``the loop'' such that certain actions and consequences must require human involvement and culpability.
As technology advances and the need for autonomous systems to handle high-variability environments increases, the relationship between the human operator and advanced AI will continue to be a critical focus.
However, it is worth commenting that these systems are not dependent on development of artificial \textit{general} intelligence, which is an orthogonal goal.
Beyond weapon systems, thrusts into neurotechnology are discussed, along with the concept of the third offset strategy, which is about advances in ``human-machine collaborative combat networks.''
A key point that is discussed is the importance of how these systems and approaches will be governed and regulated in the future, at the international level.
The discussion that followed addressed concerns about focusing on weapon systems, and elaborated on the shared goals across research and the military to use technology to avoid conflict in the first place.

\paragraph{Robin Zebrowski: Mind-body problem proliferation: The more we learn, the less we know}
This presentation is broken into two parts.
First, a dialogue and discussion is led as a plea to motivate interdisciplinary work.
This plea is premised on the argument that questions about consciousness are fundamentally dependent on philosophy, while philosophy is incapable of answering said questions, thus necessitating cross-discipline collaboration.
The second portion of the talk discusses how embodiment is fundamental to discussions of consciousness, evidenced by tremendous interdisciplinary reports that consciousness is deeply dependent on the embodiment of the agent that may be experiencing said consciousness.
Of course, embodiment is not a simple answer, as it leads to complications stemming from a lack of formal definitions and conceptualizations.
Thus, this warrants further cross-disciplinary discussion and research that respects the various fields of study while also acknowledging their limitations.

\paragraph{David Gamez: The scientific study of consciousness}
This talk explicitly addresses the four motivating questions behind this workshop.
Firstly, what is consciousness?
Consciousness, as it is is typically considered, can frequently be labeled as \textit{na{\"i}ve realism} insofar as it describes the experience of the physical world, and disregards the complex invisible factors that constitute reality (e.g., atoms and energy outside our physical sensor system).
Thus, we can distinguish consciousness as the \textit{secondary qualities}, or the subjective experience, of \textit{primary qualities}, which are all of the properties of the physical world that go beyond our sensing capabilities.
Towards understanding the mechanistic underpinnings of consciousness, it is necessary to develop a theory (\textit{C-theory}) that links the spatiotemporal patterns of physical materials (such as neurons) to conscious states.

When attempting to develop metrics of consciousness, one should distinguish between attempts to measure, predict, or deduce consciousness.
In sum, per the author: ``We \textit{measure} consciousness through first-person reports from systems that are assumed to be conscious and capable of making accurate reports about their consciousness.
We use mathematical c-theories to make testable \textit{predictions} about consciousness on systems that make reliable first-person reports.
We can make untestable \textit{deductions} about the consciousness of systems who cannot make reliable first-person reports.''
Towards machine consciousness, the definitions of four gradations of machine consciousness that were presented during the Focused Session 6 (\textit{From Human to Machine Consciousness}) are presented again here.

This talk ends with a ``plea for sanity'' that contrasts the fear of autonomous killer robots with the reality of AI capabilities, and raises the point that humans, not machines, have been extremely dangerous and have developed highly lethal devices to be used on each other for a long time.
In other words; ``People kill each other a lot. This is not going to stop.''
This suggests we should focus on making machines safe and consider ways they can reduce harm, rather than worry about a speculative future where they may be more destructive than humans.

\paragraph{Christof Koch: The Neural Correlates of Consciousness and its Implications for Architectures Supporting Consciousness}
This presentation is an abbreviated version of the previous talk that was given during Focused Session 3, \textit{Neural Correlates of Consciousness - Progress and Problems}.
The discussion that followed explored ideas of contentless consciousness.
Frequently conscious experience is framed based on the contents that are being attended to by the observer (visual perception, for instance), so the capability to determine the neural correlates of contentless consciousness may reveal a ``pure'' form of consciousness.
The presentation and discussion were focused on a firm argument that some physical property and material must be responsible for the experience and manifestation of consciousness, independent of whether we have yet appropriately developed tools that can measure said physical property.

\paragraph{John Murray: Other Voices}
Various opinions and thoughts were presented, taken from individuals who could not be present for this particular workshop session.
\textit{Adam Russell:} Collective consciousness, such that consciousness is a relative property based on what one is conscious of, is able to be shaped and manipulated by digital content. Is now the best and most critical time to best quantify and understand collective consciousness?
Directions on researching human-robot interactions and social constructs were presented, based on input from \textit{Johanna Seibt (Aarhus University)}, \textit{Thomas Sheridan (MIT)}, and \textit{Wendy Ju (Stanford University)}.
Regarding human's more precarious relationship with future autonomous systems, concerns and directions related to autonomous weapon systems from \textit{Zac Cogley (Northern Michigan University)} were presented.
Lastly, theories and potential mechanisms of consciousness from \textit{David Rosenthal (City University of New York)} and \textit{Bernard Baars (The Neurosciences Institute)} were presented, leading to the conclusion that more work is needed!

\paragraph{Joanna Bryson: A Role for Consciousness in Action Selection}
This talk provides an in-depth overview of action selection and a theoretical function of consciousness in dealing with action selection.
It covers a range of topics related to memory, biases, action, learning, and ethics.
The presentation elaborates on a previous publication \cite{bryson_role_2012}, whose abstract is included here;
``This article argues that conscious attention exists not so much for selecting an immediate action as for using the current task to focus specialized learning for the action-selection mechanism(s) and predictive models on tasks and environmental contingencies likely to affect the conscious agent. It is perfectly possible to build this sort of a system into machine intelligence, but it would not be strictly necessary unless the intelligence needs to learn and is resource-bounded with respect to the rate of learning versus the rate of relevant environmental change.
Support for this theory is drawn from scientific research and AI simulations.
Consequences are discussed with respect to self-consciousness and ethical obligations to and for AI.''
The talk combines previous concepts from the aforementioned publication, and merges it with a roadmap for conscious machines \cite{ArrabalesLS09}, with a focus on the how to best ensure optimal collaboration between humans and future AI.

\paragraph{Shannon Vallor: Overcoming barriers to cross-disciplinary consciousness research}
Provided the multifaceted complexity of consciousness research, it is a nature fit for cross-disciplinary (CD) research.
While CD research is laudable, even when the right disciplines come together, it can be fraught with complications that lead to subsequent failure.
This talk presents a guideline for successful CD research, from aligning incentives to overcoming terminological confusion.
CD research is not naturally aligned with a number of external incentives, as domain-specific expertise is typically prized based on funding, publications, and prestige.
Thus, for these types of collaborations to succeed, they must have clearly defined and alignable goals, an intrinsic desire and motivation to work together, and the appropriate amount of expertise-overlap to allow for across-discipline conversations while also affording serendipitous discovery and exploration.
However, if collaborators can overcome ``purity cultures'' that prize domain-specificity, misaligned goals, and resource problems, it is possible to develop fruitful CD research that cultivates interactional expertise.

\subsubsection{Micro-Presentations}
The micro-presentations afforded the attendees an opportunity to provide a short presentation on a research project or idea related to the workshop agenda.
Most of the attendees provided presentations, which have been summarized, while others provided abstracts or text that is presented here.
Text that is provided by the authors is clearly marked as being \textit{author-supplied} to distinguish their content from the presentation summaries that were generated by SRI.

\paragraph{Christof Koch: Responsibility and Consciousness}
In collaboration with Dr. Giulio Tononi, this project aims to understand causal responsibility, or free will.
Following from Integrated Information Theory (IIT), measures of $\phi$ will be used in humans and simulated agents to isolate extrinsic and intrinsic forces that led to a behavior or decision.
By doing so, it may be possible to isolate whether causes were based on external factors, or internal factors that can be attributed to a sense of free will.
He also proposes to ``measure the neural correlates of free will in humans (normal volunteers, impaired, children) using fMRI and EEG.''
This work would have impact on understanding and assigning responsibility in autonomous agents, both biological and artificial.

\paragraph{Henk Barendregt: Discreteness of consciousness and its consequences}
During Focused Session 4 (see slides for further detail), Henk Barendregt refined a formal model of phenomenal consciousness with input from several attendees.
This proposal is to evaluate the fear that stems from the appreciation, in non-advanced meditators, of the fleeting nature of a mutable and illusionistic “self” over which we exert no free will.
He wishes to investigate the mechanisms of overcoming such fear through specific techniques and training in mindfulness meditation.
Mindfulness meditation may have a role in the treatment of psychiatric disorders.

\paragraph{David Sahner: Evolution in the hive}
The objective of this research is to ``evaluate whether evolution of consciousness in artificial complex systems promotes moral behavior and enhances functionality.''
The approach, taking insight from modern theories of consciousness, is to use neuromorphic architectures and evolutionary algorithms to create and study embodied agents in a social context.
Phronesis will be a controlled parameter that will be studied across conditions, and frequent testing for consciousness will be used to evaluate for the presence of particular capabilities and to understand functional impact.
In particular, capabilities such as morally-guided behavior (altruism, socially-positive interactions) will be assessed, along with various tests of aspects of consciousness such as sensorimotor capabilities and $\phi$.

\paragraph{Subhash Kak: Investigating ``Consciousness Capacities'' in a Quantum Learning System}
\textit{Author-Supplied Abstract):}
``Quantum mechanics is the most fundamental theory in physics which almost certainly plays a role in the processing in the brain.
Furthermore, since quantum computing is more powerful than classical computing, one needs to go beyond the classical computation regime of present-day machines to better emulate the brain.

The field of quantum decision theory is being increasingly applied to real world problems.
Based on the mathematics of Hilbert spaces, this formalism has the capacity to capture aspects of cognitive and social processing that are not accessible to classical models.
Furthermore, it has the potential to find relationships and underlying structure in large data sets that might otherwise not be computationally feasible.
It is known that quantum models describe physical data in many aspects of biology such as photosynthesis, magnetoreception in birds, and the olfactory system, to name just a few.
Therefore, their applicability to real world data cannot be ruled out.
The proposed research will investigate the theory and implementation of quantum cognitive machines that can learn as a step in the development of recursive and self-organizing structures that is characteristic of biological systems. Consideration of quantum resources such as superposition and entanglement will open up new possibilities in creating capacities that approach consciousness.
We will also investigate quantum decision theory to the problem of choosing between alternatives with different payoff under the conditions of imperfect recall and varying degrees of knowledge of the system.
The classical version of the problem has bearing on the general area of rational agency, learning, social choice, mechanism design, auctions, and theories of knowledge.

There are results that agents with access to quantum resources can obtain superior performance as compared to classical agents. It can also be shown that in the general case where each node can have any branching probability value, a random tree can accurately represent the probabilities associated with an arbitrary quantum state.
The research on quantum agents will help in the design of general learning and decision systems that have the potential of developing capacities (levels of consciousness) that go beyond present-day AI methods.

Quantum associative memory is exponential in the number of qubits (unlike the linear capacity of classical memory).
We would like to generalize this into the development of quantum pattern classification networks quite like the instantaneously trained networks of classical computing.

\begin{figure}[h]
  \begin{center}
    \includegraphics[width=.6\linewidth]{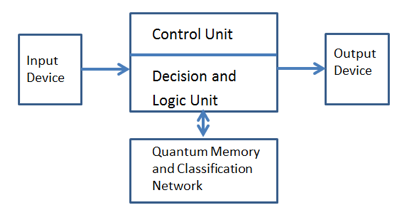}
    \caption{  \label{fig:kak}A cognitive architecture using quantum resources.}
  \end{center}

\end{figure}

The research in the project will consider the following specific problems:
Problem 1: For real decision trees from an application area, find method to determine the best quantum-equivalent representation and algorithms for efficient decisions.
Problem 2: Develop a solution to quantum pattern classification networks that learn instantaneously (which would be a generalization of quantum associative memory model).
Problem 3: Investigate requirements associated with the implementation of the cognitive architecture shown in Figure \ref{fig:kak}.

\paragraph{John Rushby: Emergence [and Consciousness]: Report of a Meeting}
We often hear of consciousness ``emerging'' from the brain, but what does this term mean?
I will briefly report on a workshop I attended on the topic of emergence (in distributed systems) organized by Herman Kopetz of TU Vienna in March 2016.
In addition to computer scientists, the participants included philosophers, biologists and others.

The consensus that emerged (!) is that we could all subscribe to what philosophers call ``weak'' emergence.
A phenomenon at the macro-level is \emph{emergent} if and only if it is \emph{of a new kind} with respect to the phenomena of its micro level parts.
An example is temperature and pressure in a gas (macro), vs.\ the motion of its molecules (micro).
The macro behavior is explainable (in principle---we may currently lack the knowledge to do so, as is the case with consciousness) in terms of the micro behavior.
I propose research to develop a more comprehensive theory of weak emergence, both as a design tool, and as a means of analysis (to be applied to consciousness).

\paragraph{Antonio Chella: Phenomenal and Social Aspects of Consciousness by Developmental Robotics}
\textit{Author-Supplied Abstract:} ``The idea for future research in consciousness studies concerns the investigation of the intertwining between phenomenal consciousness and social aspects of consciousness in the development of human beings.
The central hypothesis, inspired by “I and Thou” by the philosopher of religion Martin Buber, is that the  I-consciousness, the It-consciousness and Thou-consciousness evolve at the same time during the development of the human being.
Therefore, phenomenal consciousness is related to the perception of the external world, i.e., the relation I – It, and also to the development of the social aspects of consciousness, i.e., the relationship I - Thou, as the capability to develop a theory of other minds.
In turns, this ability is endowed with self-consciousness, i.e., the capability to develop a theory of own mind. Eventually, the ability to report phenomenal states would arise.
The proposed research will take into account the methodologies of developmental robotics to build an evolving robot child able to perceive the external world and to develop a simple theory of other minds.
The central goal of the research is to analyze whether the evolving robot may be capable of developing at the same time some form of self-consciousness and reporting phenomenal states.
A starting point will be the theory of social communication as a meeting of minds proposed by Peter G{\"a}rdenfors.
Collaborations will be placed with John Rushby, SRI to investigate the relationships of the proposal with the HOT Theories of Consciousness; with Angelo Cangelosi, Plymouth University, UK, a leading expert in developmental robotics; and with Peter G{\"a}rdenfors, Lund University, Sweden concerning the meeting of minds hypothesis.''

\paragraph{Owen Holland: The engineers of souls}
\textit{Author-Supplied Abstract:}
``While the philosophical and psychological interest in machine consciousness may be focused on phenomenal consciousness, the likely applications will primarily exploit the functional aspects of cognition that are uniquely enabled by consciousness, or by the cognitive components that necessarily underpin consciousness. Candidate elements include but are not limited to: functional awareness of an embodied and narrative self, reconstructive episodic memory, imagination, evaluation of the consequences of real or hypothetical events, planning, reflection, rationality, awareness of other minds, communication with other minds, and collaboration. In order to study some of these, which may be strongly linked to or inseparable from the others, we need multiple agents in a rich and dynamic world. Experience tells us that using multiple robots in the real world is too slow and inflexible for carrying out the necessary variety of experiments, and so we need to move to virtually embodied agents in a virtual world. This allows multiple parallel instantiations running faster than real time, with complete access to all historical states and processes in both agents and environment. In order to bring the system into the real world, the virtual system will be physics based, the environment will be an accurate capture of the real world, and the virtual robots will be accurate physics based copies of carefully designed or chosen real robots with exactly the same controllers as the virtual robots, equipped with methods for recording all internal states and processes. In principle, an experiment run in the virtual world, perhaps involving controllers trained or evolved over millions of previous experiments, could then be repeated in the real world. It would also be possible for a human to control an avatar or avatars in the virtual world in the form of the agents under study, and to control a real robot avatar in the real world. The relevant techniques for doing all this already exist to varying degrees - for example, check out the technologies underpinning Facebook's virtual, augmented, and mixed reality.

The cognitive architecture of the agents/robots should be based on what is now known as predictive processing, as this is the only prospective technology capable of spanning the range of abilities required for rational perception and action combined with the various cognitive features associated with consciousness. At present, it has two serious disadvantages: it is being oversold, and it is not as yet a usable technology. It is being oversold by claims such as: it is what all intelligent systems must do, that it is all that they must do, and that the structure of the brain has evolved to provide a recognisable implementation of it. For our purposes, we do not need a credible pseudo-neural implementation, nor do we need any credible evolutionary path for producing it, and we will be happy to supplement the predictive processing with existing technologies and architectural components as required. It is not yet usable because the philosophers and neuroscientists promoting it - with good reason - are not engineers. We are.

Although the first part of this proposal is clearly independent of the second, and is capable of supporting many other investigations, the second is clearly dependent on the first, and that is why I have presented them both together. The first is essentially an engineering development activity that could start tomorrow, but the second requires a very substantial joint scientific and technological effort from engineers, most of whom will start out knowing too little about consciousness, and psychologists, neuroscientists, and philosophers, most of whom will start out knowing too little about engineering. And though the aim is to discover and delineate the relevance of consciousness for cognitive functions, who knows where it will end up? Joseph Stalin is on record as remarking 'Writers - the engineers of the human soul' but that was then, and this is now.''

\paragraph{Joanna Bryson: Three Project Areas}
Three research agendas were proposed.
First; a project to understand the interactions between consciousness, performance, and learning.
This aims to understand how awareness impacts the expression of pre-existing knowledge and new learning.
Second; understand how implicit and explicit beliefs interact with identity, trust, and cooperation.
Bias impacts human behavior, so it would be useful to understand how group identity and polarization are impacted by conscious and non-conscious processing.
Third; ``Create sensible metaphors; processes; state, national \& transnational organisations to encourage appropriate incorporation of AI into human society.''
Because humans are the responsible agents creating robotic artifacts, their machine nature should be transparent and appropriate governance needs to be established.

\paragraph{Naveen Sundar Govindarajulu}
Proving a machine is consciousness may be impossible, so it is reasonable to argue that a machine that behaves and functions like a conscious agent is a reasonable goal for artificial intelligence.
Towards increasing ``cognitive consciousness'' in AI, it is argued to focus on reasoning, planning, and human learning (or, \textbf{G}ood \textbf{O}ld \textbf{F}ashioned \textbf{AI}), to work towards human-level cognitive behaviors.
One direction to pursue would be using theorem provers and to aim for accomplishing Theory of Mind.

\paragraph{David Gamez: Research projects on consciousness}
David Gamez proposed a number of research directions based on his gradations of machine consciousness, which he presented during Focused Session 6.
Singularity Machines; although potentially impossible, the attempt to build a machine that can build an even more intelligent machine, ad infinitum, in order to achieve the singularity would likely result in a deeper understanding and exploration of intelligence in machines.
Global Workspace; merging deep learning architectures with global workspace architectures could lead to more explainable AI and lead to machines that think more like humans.
Imagination; develop robots that use simulation and imagination so that AI can predict or ``think'' about the future and what to do next, to get machines to think in a human-like way.
Lastly, Brain Uploading; attempt to build a functionally-accurate real-time brain simulation and then explore novel attempts to measure machine consciousness ($\phi$ or examine memory space) and hardware architectures that might better support the simulation (e.g., neuromorphic chips).

\paragraph{John Sullins: Artificial Phronesis}
Phronesis is defined as ``the habit of making the right decisions and taking the right actions in context, and relentless pursuit of excellence for the common good.''
Human phronesis depends upon its evolution in a social context through learning.
Autonomy in the absence of such ethical grounding poses a danger.
This project proposes to work on developing artificial phronesis in AI agents to promote ethical behaviors.
Artificial phronesis aims to enable an agent to identify ethical situations, attend to them, and then practically and efficiently react with justified actions/decisions.

\paragraph{Julia Mossbridge: Toxic Individuality}
This project addresses the concept of toxic individuality; a concept analogous to toxic masculinity whereby there is a need to be ``in the limelight'' and there is a negative (or violent) reaction when this need is not met.
The proposal is to improve the ego development in the dominant group (males) to improve their sense of belonging and to allow them to be open to bringing others (``the enemy,'' outgroups) into their dominant group.
To do this, an AI personal coach will be used (``Mensch Mentors'').
In a pilot population of male employees, these mentors will be used to query stimulating narrative about future goals and to support and challenge individuals in a timely manner.
It is expected that a group of individuals who have these mentors will have increased well-being and ego development that will influence their relationships and subsequently enhance interpersonal and group dynamics.

\paragraph{Ian Horswill: Modeling mammalian neuropsychology for interactive virtual characters}
Humans are social animals, who often rely on emotions to guide behaviors.
While we appreciate this component of ``humanity,'' we typically disregard complex social and emotional components of human behavior when we start to talk about advanced AI.
Rather than focus simply on the human-level cognitive component when building AI, it is argued here that we should consider the more social and emotional components of agent behavior.
This project aims to build a model of a social mammal to observe and learn from it.
Towards understanding emotions, it is conjectured that some emotions are just activations states of the behavioral system with which they are associated.
Thus, emotions may merely the outward behavior of an underlying algorithm or system activation, rather than a stand-alone process that modulates or motivates other behaviors.

\subsection{Workshop Attendees}
\label{sec:attendees}

The conference was attended by thirty-two invited attendees from a
variety of international external institutions, and sixteen
attendees from SRI.  The
full attendee list of the 2017 SRI Technology \& Consciousness
Workshop Series follows;

\begin{itemize}
\item[]
\textbf{Last Name, First Name}
\textit{Affiliation}\\
Research Interests and Expertise
\end{itemize}

\paragraph*{External Attendees}
\begin{itemize}
\item[]
\textbf{Barendregt, Henk}
\textit{Nijmegen University; Chair, Foundations of Mathematics and Computer Science}\\
Lambda calculus and type theory; mathematical definition of mindfulness; mindfulness meditation
\item[]
\textbf{Bickhard, Mark}
\textit{Lehigh University; Henry R. Luce Professor of Cognitive Robotics and the Philosophy of Knowledge}\\
Cognitive robotics, philosophy of knowledge, interested in theoretical physics and mathematics
\item[]
\textbf{Blackmore, Susan}
\textit{Visiting Professor in Psychology at the University of Plymouth}\\
Widely recognized expert in consciousness studies and author of textbook on this topic and other relevant books
\item[]
\textbf{Bringsjord, Selmer}
\textit{Director, Rensselaer Polytechnic Inst. AI \& Reasoning Laboratory; Chair \& Professor of Cognitive Science; Professor of Computer Science; Professor of Logic \& Philosophy}\\
AI and cognitive science, philosophy of mind
\item[]
\textbf{Bryson, Joanna}
\textit{University of Bath}\\
Artificial models of natural intelligence
\item[]
\textbf{Chalmers, David}
\textit{Director, Center for Mind, Brain and Consciousness, New York University}\\
Renowned philosopher who has contributed greatly to the understanding of human consciousness
\item[]
\textbf{Chella, Antonio}
\textit{University of Palermo}\\
Machine consciousness, robotics, computer vision
\item[]
\textbf{Earth and Fire Erowid}
\textit{Erowid Center}\\
Maintain extensive database of human experiences with psychoactives
\item[]
\textbf{Gamez, David}
\textit{Middlesex University Department of Computer Science}\\
Measurement of consciousness; information integration theory of consciousness
\item[]
\textbf{Govindarajulu, Naveen Sundar}
\textit{Rensselaer Polytechnic Institute, Troy, New York}
Data Scientist; natural language processing
\item[]
\textbf{Grabowecky, Marcia}
\textit{Northwestern University}\\
Multi-sensory integration, perception, and meditation
\item[]
\textbf{Holland, Owen}
\textit{University of Sussex Sackler Center for Consciousness Science}\\
Machine consciousness and cognitive robotics
\item[]
\textbf{Horswill, Ian}
\textit{Northwestern University}\\
AI, control systems based on goal state and sensor
\item[]
\textbf{Kaiser-Greenland, Susan}
\textit{Mindfulness educator and author}\\
Mindfulness
\item[]
\textbf{Kak, Subash}
\textit{Oklahoma State University School of Electrical and Computer Engineering}\\
Quantum cognitive science, neural networks, and quantum computing
\item[]
\textbf{Koch, Christof}
\textit{President/CSO, Allen Institute for Brain Science}\\
Cognitive and behavioral biology, artificial intelligence; theoretical, computational and experimental neuroscience, consciousness studies
\item[]
\textbf{Moreno, Jonathan}
\textit{University of Pennsylvania}\\
History of science and medicine, psychology, bioethics, served on several NAS and DoD studies regarding military hardware, autonomy, ethics
\item[]
\textbf{Mossbridge, Julia}
\textit{Northwestern University and Institute of Noetic Sciences (IONS)}\\
Duality, consciousness, predictive anticipatory activity
\item[]
\textbf{No{\"e}, Alva}
\textit{Professor of philosophy at the University of California, Berkeley; member of the Institute for Cognitive and Brain Sciences and the Center for New Media}\\
Perception and consciousness, theory of art, phenomenology, extended mind and embodiment
\item[]
\textbf{Presti, David}
\textit{University of California, Berkeley}\\
Neurobiologist and cognitive scientist
\item[]
\textbf{Rensink, Ron}
\textit{University of British Columbia}\\
Vision systems and computer science
\item[]
\textbf{Rosenthal, David}
\textit{Professor of Philosophy; Coordinator, Interdisciplinary Concentration in Cognitive Science, CUNY}\\
Philosophy of Mind, metaphysics
\item[]
\textbf{Rowe, Bill}
\textit{Formerly of Santa Cruz Institute for Particle Physics; consultant to neuroscience companies, brain implants for preclinical studies}\\
Authority on Julian Jaynes, interests in language acquisition and ontogeny of consciousness
\item[]
\textbf{Schneider, Susan}
\textit{Dept. of Philosophy, Cognitive Science Program, University of Connecticut; Yale Interdisciplinary Center for Bioethics; Center of Theological Inquiry, Princeton}\\
Philosophy of Mind, ethics, neuroscience and AI, metaphysics
\item[]
\textbf{Small, Gary}
\textit{Professor of Psychiatry and Biobehavioral Sciences and Parlow-Solomon Professor on Aging at the David Geffen School of Medicine at UCLA}\\
Aging/Alzheimer’s Disease expert
\item[]
\textbf{Syverson, Paul}
\textit{Naval Research Laboratory}\\
Mathematician and computer scientist
\item[]
\textbf{Sullins, John}
\textit{Sonoma State University}
Philosophy, AI ethics, intelligent moral agents
\item[]
\textbf{Tononi, Guilio}
\textit{Professor, University of Wisconsin Madison; Center for Sleep and Consciousness}\\
Information theory of consciousness
\item[]
\textbf{Vallor, Shannon}
\textit{Santa Clara University}\\
Philosophy of Science and Technology; ethics of emerging technologies
\item[]
\textbf{Zebrowski, Robin}
\textit{Beloit College}\\
Interdisciplinary: cognitive science, philosophy, psychology, and computer science. Human concepts of embodiment, with applications to artificial intelligence and robotics. Metaphor theory.
\end{itemize}

\paragraph*{SRI Attendees and Relevant Interests}
\begin{itemize}
  \item[]
  \textbf{Lincoln, Patrick}
  \textit{Director, Computer Science Lab}\\
  {[Principal Investigator]} Formal methods and logic, computational biology, interest in human consciousness
  \item[]
  \textbf{Sahner, David}
  \textit{Senior Clinical Director, Computer Science Lab}\\
  {[Technical Lead]} Artificial intelligence, philosophy of mind, neuroscience and cognition, medicine, and contemporary physics
  \item[]
  \textbf{Murray, John}
  \textit{Program Director, Computer Science Lab}\\
  {[Technical Co-lead]} Neuroergonomics, interactive collaboration in real and virtual environments, cognitive engineering, and cyber-research ethics
  \item[]
  \textbf{Adkins, Boone}
  \textit{Software Engineer, Advanced Technology and Systems Development}
  \item[]
  \textbf{Atrash, Amin}
  \textit{Software Engineer, Advanced Technology and Systems Development}\\
  robotics, autism
  \item[]
  \textbf{Connolly, Christopher}
  \textit{Principal Scientist, Computer Science Lab}\\
  artificial intelligence, computing in mathematics, engineering, medicine, and neuroscience; neurogram deconvolution
  \item[]
  \textbf{Israel, David}
  \textit{Principal Scientist Emeritus, Artificial Intelligence Lab}\\
  AI, natural language representation and reasoning
  \item[]
  \textbf{Keifer, Kellie}
  \textit{Senior Operations Director, Computer Science Lab}
  \item[]
  \textbf{Kirkwood, Gordon}
  \textit{Research Engineer, Advanced Technology and Systems Development}
  \item[]
  \textbf{Myers, Karen}
  \textit{Program Director, Artificial Intelligence Lab}
  \item[]
  \textbf{Poggio, Andy}
  \textit{Senior Computer Scientist, Computer Science Lab}\\
  computing in natural sciences, mathematics, and a clinical and biological context
  \item[]
  \textbf{Rushby, John}
  \textit{Principal Scientist, Computer Science Lab}\\
  humans and safety-critical systems, theory of mind
  \item[]
  \textbf{Sanchez, Daniel}
  \textit{Cognitive Scientist, Computer Science Lab}
  applied cognitive neuroscience, human learning and memory, human-intuitive machines
  \item[]
  \textbf{Shankar, Natarajan}
  \textit{Principal Scientist, Computer Science Lab}\\
  fundamental mathematics, foundational logic, software verification
  \item[]
  \textbf{Willoughby, Adrian}
  \textit{Cognitive Scientist, Computer Science Lab}
  \item[]
  \textbf{Williams, Damien}
  \textit{Scientific Conference Assistant, Computer Science Lab}
\end{itemize}

\end{document}